\documentclass{jfm}
\usepackage{graphicx}
\usepackage{epstopdf, epsfig}

\usepackage{natbib}
\usepackage{graphicx}
\usepackage{float}
\usepackage{subfigure}
\usepackage{amsmath}
\usepackage{amssymb}
\usepackage{color}
\usepackage{soul}

\graphicspath{{Figures/}}

\newcommand{\ii}[0]{\mathrm{i}}

\newcommand{\ee}[0]{\mathrm{e}}

\shorttitle{Absolute instability in shock-containing jets}
\shortauthor{P. A. S. Nogueira et al.}

\title{Absolute instability in shock-containing jets}

\author{Petr\^onio A. S. Nogueira\aff{1}
  \corresp{\email{petronio.nogueira@monash.edu}},
  Peter Jordan\aff{2},
  Vincent Jaunet\aff{2},
  Andr\'e V. G. Cavalieri\aff{3},
  Aaron Towne\aff{4},
  \and
  Daniel Edgington-Mitchell \aff{1}}

\affiliation{\aff{1}Department of Mechanical and Aerospace Engineering, Laboratory for Turbulence Research in Aerospace and Combustion, Monash University, Clayton, Australia
\aff{2}Département Fluides, Thermique, Combustion, Institut PPrime, CNRS–Université de Poitiers–ENSMA, Poitiers, France
\aff{3} Divis\~ao de Engenharia Aeron\'autica, Instituto Tecnol\'ogico de Aeron\'autica, S\~ao Jos\'e dos Campos, SP, 12228-900, Brazil
\aff{4}Department of Mechanical Engineering, University of Michigan, 2350 Hayward Street, Ann Arbor, MI 48109, USA}

\begin{document}

\maketitle

\begin{abstract}
We present an analysis of the linear stability characteristics of shock-containing jets. The flow is linearised around a spatially periodic mean, which acts as a surrogate for a mean flow with a shock-cell structure, leading to a set of partial differential equations with periodic coefficients in space. Disturbances are written using the Floquet ansatz and Fourier modes in the streamwise direction, leading to an eigenvalue problem for the Floquet exponent. The characteristics of the solution are directly compared to the locally parallel case, and some of the features are similar. The inclusion of periodicity induces minor changes in the growth rate and phase velocity of the relevant modes for small shock amplitudes. On the other hand, the eigenfunctions are now subject to modulation related to the periodicity of the flow. Analysis of the spatio-temporal growth rates led to the identification of a saddle point between the Kelvin-Helmholtz mode and the guided jet mode, characterising an absolute instability mechanism. Frequencies and mode shapes related to the saddle points for two conditions (associated with axisymmetric and helical modes) are compared with screech frequencies and the most energetic coherent structures of screeching jets, resulting in a good agreement for both. The analysis shows that a periodic shock-cell structure has an impulse response that grows upstream, leading to oscillator behaviour. The results suggest that screech can occur in the absence of a nozzle, and that the upstream reflection condition is not essential for screech frequency selection. Connections to previous models are also discussed.
\end{abstract}

\begin{keywords}
Authors should not enter keywords on the manuscript.
\end{keywords}

\section{Introduction}
\label{sec:intro}


With the development of commercial and military aviation in the last century, the importance of studies on noise generation by jets has increased substantially. These studies can be divided into two broad categories, based on the source mechanisms responsible for noise generation. The first category is shock-free jets, comprising both subsonic jets and ideally-expanded supersonic jets; the acoustic field of such flows is usually dominated by the noise generated by organised structures present in the turbulent flow \citep{jordan2013wave}, which underpin both Mach wave radiation (for supersonic jets), and turbulent mixing noise (subsonic and supersonic cases). The second category is shock-containing jets, i.e., imperfectly expanded supersonic jets. In this case, the pressure deficit between the choked nozzle and external medium leads to the appearance of a spatially periodic coherent structure formed by several shock-cells \citep{pack1950}. The appearance of shock cells dramatically impacts the acoustics of these jets, with the appearance of two other components due to the interaction between the organised structures and the shocks \citep{Tam1995}: broadband shock-associated noise (BBSAN), usually peaking at slightly higher frequencies than the turbulent mixing noise, and screech tones, associated with high intensity pressure fluctuations at specific frequencies. 

The earliest description of the screech phenomenon was that of \cite{powell1953}. Using schlieren photographs, Powell identified the presence of large-scale turbulent structures travelling downstream and acoustic waves travelling upstream in shock-containing jets. Motivated by these observations, he proposed that screech tones are generated by a self-sustained resonant process involving both waves, where the interaction between the large-scale structures and the shocks gives rise to acoustic waves that excite new downstream-travelling structures upon reaching the nozzle lip. As usual in the analysis of resonant phenomena, both phase and gain conditions for this process were derived \citep{Powell_1953_On} and applied to the study of screeching jets. Powell's theory managed to predict screech tones and directivities with relative success, being used as the foundation of several subsequent analyses, as summarised by \cite{RAMAN1998} and \cite{EdgingtonMitchell2019}.

While the downstream propagation of energy is quite well understood, the process of upstream wave generation in this cycle is still under investigation. While the original work of \cite{powell1953} proposes a multiple, discrete-source formulation, related to the position of each shock-cell, other works have proposed alternative views. For instance, \cite{TamTanna1982} proposed a formulation based on a frequency-wavenumber analysis of the different waves in the jet, leading to a continuous source distribution in the turbulent medium. In this previous work (also in \cite{TamSeiner1986}), the authors propose that screech is a direct consequence of the interaction between shock-cell and large-scale structures in the wavenumber domain. In this framework, no position for the generation of upstream waves is imposed, and predictions are performed by analysing the wavenumbers energised by such interaction. More recently, \cite{gojon2018aiaa} and \cite{edgington-mitchell_jaunet_jordan_towne_soria_honnery_2018} have shown that the resonance mechanism may actually be closed by a neutrally stable guided jet mode, as opposed to an acoustic wave. This mode has specific bands of existence that roughly match the regions where screech tones are found experimentally. Resonance models using these guided jet modes lead to a better alignment with experimental data, strongly supporting this hypothesis \citep{mancinelli2019,mancinelli2020,Nogueira2020A1A2}. The work of \cite{manning1998numerical,manning2000numerical} proposed a mechanism for generating upstream waves, in which the complex interaction between vortices and shocks would give rise to acoustic waves. The model managed to reproduce characteristics of screech observed in experiments \citep{ShariffManning2013,edgington-mitchell_weightman_lock_kirby_nair_soria_honnery_2021}, even with strong simplifications of its underlying assumptions.

While early models for the analysis of jet turbulence, including screech, were based on experimental observations and classic resonance theory, the development of stability theory helped to uncover the physics of the problem. Tools based on this theory have been used extensively in the analysis of flow dynamics and sound generation in subsonic jets \citep{michalke1964inviscid,michalke1965spatially,crow1971orderly,michalke1971instabilitat,crighton1975bpa,cavalieri2012axisymmetric,baqui2015coherence,cavalieri2019amr}. While most of these tools were developed using a locally parallel framework, which disregards the spatial evolution of the mean flow, some recent works managed to account for such effects by using global stability \citep{coenen_lesshafft_garnaud_sevilla_2017,schmidt2017} and resolvent analysis \citep{garnaud2013,jeun2016,schmidt2018spectral,towne2018spectral,lesshafft2019resolvent,pickering2019lift}. Since the spatial inhomogeneity in imperfectly expanded jets is even stronger due to the presence of the shock-cell structure, application of global methods is usually more appropriate, even though it demands a much higher computational power. Examples of analyses using such tools in shock-containing jets can be seen in \cite{BENEDDINE2015} and \cite{EdgingtonMitchell2020}. In these previous works, screech can be seen as a global mode of the flow, when the shock-containing mean flow is considered as the base flow. Even though global methods may accurately capture the screech frequency and the overall characteristics of the dominant flow structures, they do not directly reveal the physical mechanisms at play; simpler models can often be used to gain such insight. This is exemplified in the complementary works of \cite{schmidt2017} and \cite{towne_cavalieri_jordan_colonius_schmidt_jaunet_bres_2017}: while global modes are used in the first work to extract both the frequency and structure of trapped acoustic modes in a subsonic jet, the latter paper elucidates the underlying nature and behaviour of these waves by using weakly nonparallel analysis.


One way of accounting for the additional streamwise inhomogeneity of shock-containing flows without resorting to global methods is to consider the periodicity induced by the shock-cell structure directly in the mean flow, rather than treating the mean flow as locally parallel. Such methodologies are usually used in studies of secondary instability in shear flows \citep{Herbert1988,brandt_cossu_chomaz_huerre_henningson_2003}. In these cases, the periodicity allows for the use of Floquet theory, which simplifies the solution procedure of the set of partial differential equations with periodic coefficients. The concepts of convective and absolute instabilities were extended to the spatially periodic case by \cite{Brevdo1996}; as in the locally parallel case \citep{huerre1990local}, stability analysis on spatially periodic cases can lead to three different scenarios: disturbances can be exponentially damped in space and time in all directions from the source (stable); they can be exponentially amplified in a specific direction, but convected away from the source (convectively unstable); or they can be amplified in both space and time, eventually contaminating the entire flow field (absolutely unstable). \cite{Brevdo1996} explored these scenarios by solving the Ginzburg-Landau equation linearised around a periodic base flow, showing that the solution may change its stability characteristics depending on the magnitude of the coefficients related to periodicity in the equation. In real flows, absolute instability is generally restricted to hot jets and cold wakes (and some particular flow cases with specific configurations, as summarised by \cite{huerre1990local}), where the flow behaves as an oscillator, with amplified disturbances travelling both upstream and downstream. This description qualitatively matches the overall characterisation of the screech phenomenon provided by \cite{powell1953}. However, while the few extant global analyses have accurately captured the aforementioned upstream and downstream propagating waves in a single global mode (see \cite{BENEDDINE2015}, for instance), the physical mechanism underpinning the relationship between these waves cannot be directly determined from such an analysis. No link between an absolute instability mechanism and screech has been made to this date.


In this paper, we explore the formulation developed by \cite{Brevdo1996} for a locally parallel stability analysis which accounts for the spatial periodicity of the mean flow -- the spatially periodic linear stability analysis (SP-LSA). This is applied for the first time to the study of screech in shock-containing jets. The formulation goes one step further in complexity when compared to the locally parallel case, while still neglecting the spatial spread of the mean flow to permit a much faster computation than a global stability analysis. This also allows for a clearer characterisation of the stability of the flow, with an extraction of mechanisms leading to screech. The paper is organised as follows: in section \ref{sec:stability}, the spatially periodic formulation using the Floquet ansatz is detailed. The impact of the inclusion of a periodic shock-cell structure on the different waves supported by the flow and on the stability characteristics of the flow is explored in section \ref{sec:results}. A discussion about the relationship between the present results and previous analyses is performed in \S \ref{sec:relationship}, and the paper is concluded in \S \ref{sec:concl}.

\section{Stability analysis of a streamwise periodic flow}
\label{sec:stability}

The present formulation is based on the spatio-temporal linear stability analysis as developed by \cite{Briggs1964}, \cite{bers1975linear} and \cite{huerre_monkewitz_1985}. The compressible inviscid linearised Navier-Stokes equations can be written in the matrix operator form as

\begin{equation}
    \frac{\partial \mathbf{{q'}} }{\partial t} + \mathbf{L}_x \frac{\partial \mathbf{{q'}} }{\partial x} + \mathbf{L}_r \frac{\partial \mathbf{{q'}} }{\partial r} + \mathbf{L}_\theta \frac{\partial \mathbf{{q'}} }{\partial \theta}  + \mathbf{L}_0  \mathbf{{q'}} =0,
    \label{eqn:LNS0}
\end{equation}


\noindent  where the disturbance vector is given by $\mathbf{{q'}}(x,r,\theta,t)=[ \nu \ u_x \ u_r \ u_\theta \ p]^\mathrm{T}$, which includes specific volume, streamwise, radial and azimuthal velocities, and pressure. Normal modes are assumed in time and azimuth, such that $\mathbf{{q'}}$ can be written as a function of the azimuthal wavenumber $m$, the frequency $\omega$ and the spatial coordinates $(x,r)$ as

\begin{equation}
    \mathbf{{q'}}(x,r,\theta,t)=\mathbf{\hat{q}}(x,r) \mathrm{exp}(-\ii \omega t + \ii m \theta).
    \label{eqn:response1}
\end{equation}

\noindent Thus, (\ref{eqn:LNS0}) can be rewritten as

\begin{equation}
    -\ii\omega \mathbf{I} \mathbf{\hat{q}} + \mathbf{L}_x \frac{\partial \mathbf{\hat{q}} }{\partial x} + \mathbf{L}_r \frac{\partial \mathbf{\hat{q}} }{\partial r} + \ii m \mathbf{L}_\theta \mathbf{\hat{q}}  + \mathbf{L}_0  \mathbf{\hat{q}} =0,
    \label{eqn:LNS1}
\end{equation}

The operators $\mathbf{L}_x$, $\mathbf{L}_r$, $\mathbf{L}_\theta$ and $\mathbf{L}_0$ are dependent on the spatial derivatives and the mean flow quantities $\bar{\mathbf{q}}(x,r)=[\bar{\nu} \ U_x \ U_r \ U_\theta \ P]$, which are also a function of $(x,r)$. In the present study, both the radial and azimuthal components of the mean velocity are considered to be negligible. In the locally parallel case, normal modes would also be considered in the streamwise direction, and the disturbances would be written as function of the streamwise wavenumber $k$. Here, instead of considering the flow to be locally parallel, a spatial variation in the form of a sinusoidal wave is considered as an analogue for shock structures in the flow. Thus, the time-averaged streamwise velocity is considered to have a dependence in the streamwise direction as


\begin{equation}
    U_x(x,r)=U(r)\left[1+A_{sh} \cos{\left( k_{sh} x \right)}\right],
    \label{eqn:meanflow_x}
\end{equation}


\noindent where $k_{sh}=2\pi/\lambda_{sh}$ is the shock-cell wavenumber, $U(r)$ is the radial shape of the mean flow, and $A_{sh}$ is the strength of the shock-cell structure. For the present case, the temperature is obtained from a Crocco-Busemann approximation \citep{lesshafft2007linear}, pressure is obtained from a spatial integration using the streamwise momentum equation \citep{van2007evaluation}, and the specific volume is computed using the ideal gas law. All quantities are normalised by the jet diameter $D$, the ambient sound speed $c_\infty$, and the ambient density $\rho_\infty$. 

As in \cite{michalke1971instabilitat}, the radial shape of the mean flow is given by

\begin{equation}
    U(r)=M\left[0.5+0.5\mathrm{tanh}\left(0.5\left(\frac{0.5D_j}{r}-\frac{r}{0.5D_j}\right)\frac{1}{\delta}\right)\right),
    \label{eqn:UmeanRad}
\end{equation}

\noindent where $D_j$ is the ideally expanded diameter for a simple convergent nozzle \citep{Tam1995}, and $M=M_j \sqrt{T_j/T_\infty}$ is the acoustic Mach number, computed using the ideally expanded jet Mach number $M_j$ and the temperature ratio. The parameter $\delta$ defines the shear layer thickness of the jet. It is worth noting that the oscillatory part of (\ref{eqn:meanflow_x}) is similar to the one proposed by \cite{TamTanna1982} and \cite{Tam1995} for the velocity variation induced by the shock-cell structure. Equations (\ref{eqn:meanflow_x}) and (\ref{eqn:UmeanRad}) are considered as a first approximation of the shock-cell structure, retaining some of its key characteristics, such as streamwise dependence and some features of radial shape as derived by \cite{pack1950}, at least of the dominant term of the series that represents the shock-cell structure. Choice of other radial shapes are expected to lead to similar results, as the streamwise periodicity is the key element of this analysis, as will be shown in the next sections. The mean streamwise velocity (and all other mean flow quantities) have an $x$-periodicity given by

\begin{equation}
    U_x(x,r)=U_x(x+N\lambda_{sh},r),
    \label{eqn:meanflow_per}
\end{equation}

\noindent where $N$ is an integer. Thus, (\ref{eqn:LNS1}) becomes a set of partial differential equations (PDEs) with $x$-periodic coefficients. Following \cite{Herbert1988} and \cite{Brevdo1996}, such periodicity allows us to use the Floquet ansatz and consider solutions in the shape

\begin{equation}
    \mathbf{\hat{q}}(x,r) = \mathbf{\tilde{q}}(x,r)\ee^{\ii\mu x},
    \label{eqn:response_per}
\end{equation}

\noindent where $\mathbf{\tilde{q}}(x,r)$ has the same periodicity of the base flow. In this formulation $\ee^{\ii\mu \lambda_{sh}}$ is called the \textit{Floquet multiplier}, and $\mu=\mu_r+\ii \mu_i$ is the \textit{Floquet exponent}. It is straightforward to see that, for the locally parallel case, where $\mathbf{\tilde{q}}(x,r)=\mathbf{\tilde{q}}(r)$, the Floquet exponent is simply reduced to the streamwise wavenumber $k$ in the normal mode ansatz. Due to its periodicity in $x$, $\mathbf{\tilde{q}}(x,r)$ can be expanded as a Fourier series \citep{Herbert1988},

\begin{equation}
    \mathbf{\tilde{q}}(x,r)=\sum_{n=-\infty}^\infty{\mathbf{\tilde{q}}_n(r)}\ee^{\ii n k_{sh} x}.
    \label{eqn:response_fourier}
\end{equation}

\noindent Thus, solutions related to the Floquet exponent $\mu+N k_{sh}$ (with integer $N$) can be written as

\begin{equation}
    \mathbf{\hat{q}}(x,r) = \sum_{n=-\infty}^\infty\left[{\mathbf{\tilde{q}}_n(r)}\ee^{\ii n k_{sh} x}\right]\ee^{\ii\mu x}\ee^{\ii N k_{sh} x} = \sum_{n=-\infty}^\infty\left[{\mathbf{\tilde{q}}_n(r)}\ee^{\ii (n+N) k_{sh} x}\right]\ee^{\ii\mu x},
    \label{eqn:response_per_muksh1}
\end{equation}

\noindent which means that solutions related to $\mu$ and $\mu+N k_{sh}$ cannot be distinguished, as one can be obtained from the other by simply reordering the Fourier coefficients.

By substituting (\ref{eqn:response_per}) into (\ref{eqn:LNS1}) we obtain

\begin{equation}
    -\ii\omega \mathbf{I} \mathbf{\tilde{q}} + \mathbf{L}_x \left[  \frac{\partial}{\partial x} + \ii \mu  \right] \mathbf{\tilde{q}} + \mathbf{L}_r \frac{\partial \mathbf{\tilde{q}} }{\partial r} + \ii m \mathbf{L}_\theta \mathbf{\tilde{q}}  + \mathbf{L}_0  \mathbf{\tilde{q}} =0,
    \label{eqn:LNS1_per}
\end{equation}

\noindent which allows us to write an eigenvalue problem for the complex Floquet exponent,

\begin{equation}
    \mathbf{L} \mathbf{\tilde{q}} = \mathbf{L}_{\mu} \mu \mathbf{\tilde{q}}.
    \label{eqn:eigenvalueFloquet}
\end{equation}

\noindent The operators $\mathbf{L}$ and $\mathbf{L}_{\mu}$ can be found in Appendix \ref{app:linops}.

Equation \ref{eqn:eigenvalueFloquet} has exactly the same form as the locally parallel spatial stability analysis; in fact, when $\mu=k$ both analyses are identical (as the streamwise derivatives in the operators above can also be neglected in the local analysis). When $A_{sh}=0$, the solution of the eigenvalue problem will give rise to modes following the relation $\mu=k + N k_{sh}$ in the eigenspectrum. However, if $A_{sh}>0$, i.e. the flow has a spatial variation within a wavenumber length, such modulation may change both the eigenvalues (related to phase velocity and growth rate of the different waves supported by the flow), and the shapes of the modes. Similar to the locally parallel case, downstream-travelling modes with $\mu_i<0$ will be exponentially amplified in space (unstable modes), and all modes can be classified by continuation from this previous case. The distinction between downstream- ($\mu^+$) and upstream-travelling ($\mu^-$) modes can be made by using Briggs' criterion, as used for parallel base flows \citep{Briggs1964,tam_hu_1989,towne_cavalieri_jordan_colonius_schmidt_jaunet_bres_2017}.

The similarities between the present formulation and the locally parallel analysis were further explored by \cite{Brevdo1996}, who showed that the conditions for absolute instability in spatially periodic flows are similar to the locally parallel case (as summarised by \cite{huerre1990local}): the occurrence of a saddle point (related to the appearance of a double root in the complex $\mu$ plane) for positive imaginary frequency $\omega_i>0$ is a condition for absolute instability. The impulse response of the spatially periodic base flow at a fixed position $x$ has an $\exp(-i \omega_0 t)$ time dependence for large $t$, where $\omega_0=\omega_{0,r}+\ii\omega_{0,i}$ is the frequency of the saddle point. As in the locally parallel case, the two modes involved in the saddle must also move to opposite sides of the real $\mu$ axis for increasingly large $\omega_i$, which is the equivalent to the condition that the saddle should be formed by a $k^+$ and a $k^-$ mode in this previous case. The saddle with maximal $\omega_{0,i}$ satisfying this condition is the relevant one for the calculation of the impulse response, as it is the saddle point obtained with a continuous deformation of the Fourier inversion contour starting from large $\omega_i$ to ensure causality of the impulse response \citep{huerre1990local,huerre2000open}. This requirement is also equivalent to the condition that the absolute value of the Floquet multipliers should move external to the unit circle ($|\ee^{\ii\mu \lambda_{sh}}|=1$) for the first mode, and internal to this circle for the second mode, in the limit $\omega_i \to \infty$ \citep{Brevdo1996}. Essentially, writing the equations as a function of the Floquet exponents leads to a linear problem that inherits most of the characteristics of the locally parallel case \citep{BrancherChomaz1997}; for this reason, it will be analysed in the same fashion. The detection of a saddle point with $\omega_{0,i}>0$ implies that the impulse response of a spatially periodic base flow grows in both downstream and upstream directions, leading to an oscillator behaviour with frequency $\omega_{0,r}$ that spreads throughout the domain. Around a fixed $x$, the spatial structure of the impulse response at large $t$ is given by the eigenfunction $\mathbf{\hat{q}}(x,r)$ at the saddle point.

The eigenvalue problem (\ref{eqn:eigenvalueFloquet}) is solved numerically by discretising the computational field in the radial direction using Chebyshev polynomials, and in the streamwise direction using Fourier modes  \citep{weideman2000matlab}. Radial mapping and boundary conditions were implemented as in \cite{lesshafft2007linear}, and the problem was solved numerically using the Arnoldi method (\textit{eigs} in Matlab). For the cases studied herein (especially for low $A_{sh}$), a discretisation of $N_r \times N_x=80 \times 31$ in the radial and streamwise directions was shown to be sufficient to converge all the relevant modes. Computation of 400 eigenvalues using this number of points usually takes around 180 seconds on a standard laptop for each choice of parameters.

\section{Results}
\label{sec:results}


Depending on how far a jet is operated from its design condition, the shock and expansion structures within the jet core can vary significantly in strength. Far from the design condition, such as may be the case in rocket propulsion, the shock cells can be complex structures, involving normal and oblique shocks, and their associated triple points \citep{EdgingtonMitchell2014underexpanded}. Closer to the design point, where the nozzles of air-breathing engines are likely to operate and screech is more likely to occur, the shock structures are much weaker, and the compression and expansion may take place near-isentropically. In the present analysis, we focus on two cases with relatively weak shocks (nozzle pressure ratio $NPR = 2.1$ and $2.4$ for convergent nozzles). Based on experimental data at these conditions, a shock amplitude of $A_{sh}=0.02$ is selected to approximate the oscillations observed around the fourth shock cell of these jets. Such shock amplitude ensures that the entire region where the shocks are found in the flow is supersonic. As in \cite{pack1950}, the shock-cell wavelength is approximated by $\lambda_{sh}=\pi/2.4048 \sqrt{M_j-1}$, where $M_j$ is the ideally expanded jet Mach number. Previous results show that the first case ($NPR=2.1$) is dominated by an $m=0$ (A1) screeching mode \citep{edgington-mitchell_jaunet_jordan_towne_soria_honnery_2018}, and the second ($NPR=2.4$) was shown to have a competition between A2 ($m=0$) and B ($m=1$) modes \citep{li_zhang_hao_he_2020}. Thus, the azimuthal wavenumbers of the analysis were chosen as $m=0$ for the first case, and $m=1$ for the second. For the absolute stability analysis, the shear-layer thickness was chosen as $\delta=0.15$ for the $m=0$ case, and $\delta=0.25$ for the $m=1$ case. This is in line with the results of \cite{Powell1992}, who showed that screech B modes are associated with a larger jet spread angle (see also \cite{Tan2017novel}). For the analysis of the modulation by the shock-cells, $\delta=0.15$ was kept for both cases, in order to provide a fair comparison between the effects of the shocks on the waves associated with both azimuthal wavenumbers.


\subsection{Overall characteristics of the modulation}
\label{sec:modulation}

\begin{figure}
\centering
\includegraphics[clip=true, trim= 0 50 0 50, width=1\textwidth]{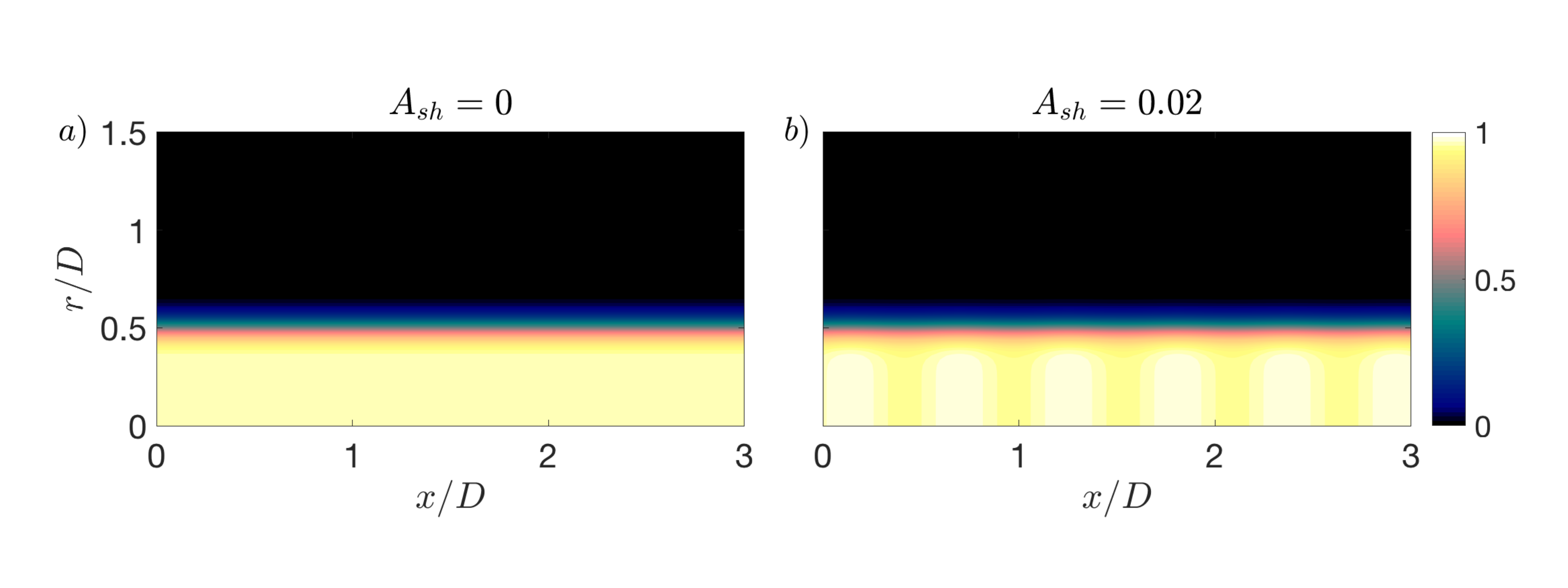}
\caption{Mean streamwise velocity for the $NPR=2.1$ case normalised by the ambient sound speed for $A_{sh}=0$ (a) and $A_{sh}=0.02$ (b).}
\label{fig:Umean}
\end{figure}

We start by analysing the effect of the periodic shock-cell structure in the different waves supported by the flow. Figure \ref{fig:Umean} shows the mean flows for $A_{sh}=0$ (equivalent to the locally parallel case), and $A_{sh}=0.02$. Figure \ref{fig:Umean}(b) exhibits some of the leading characteristics of a shock-cell structure in an under-expanded jet, though with some differences in shape when compared with experiments (see, for example, \cite{EdgingtonMitchell2020}). As noted by \cite{TamTanna1982}, the addition of other shock-cell wavenumbers may lead to a closer agreement in the shape of this structure with experiments; here, we consider the leading wavenumber to be sufficient to analyse the effect of such structure on the different waves in the flow. The only streamwise variations allowed in the shear-layer are those arising from the presence of the shock-cell structure in the flow.

\begin{figure}
\centering
\includegraphics[clip=true, trim= 0 0 0 0, width=\textwidth]{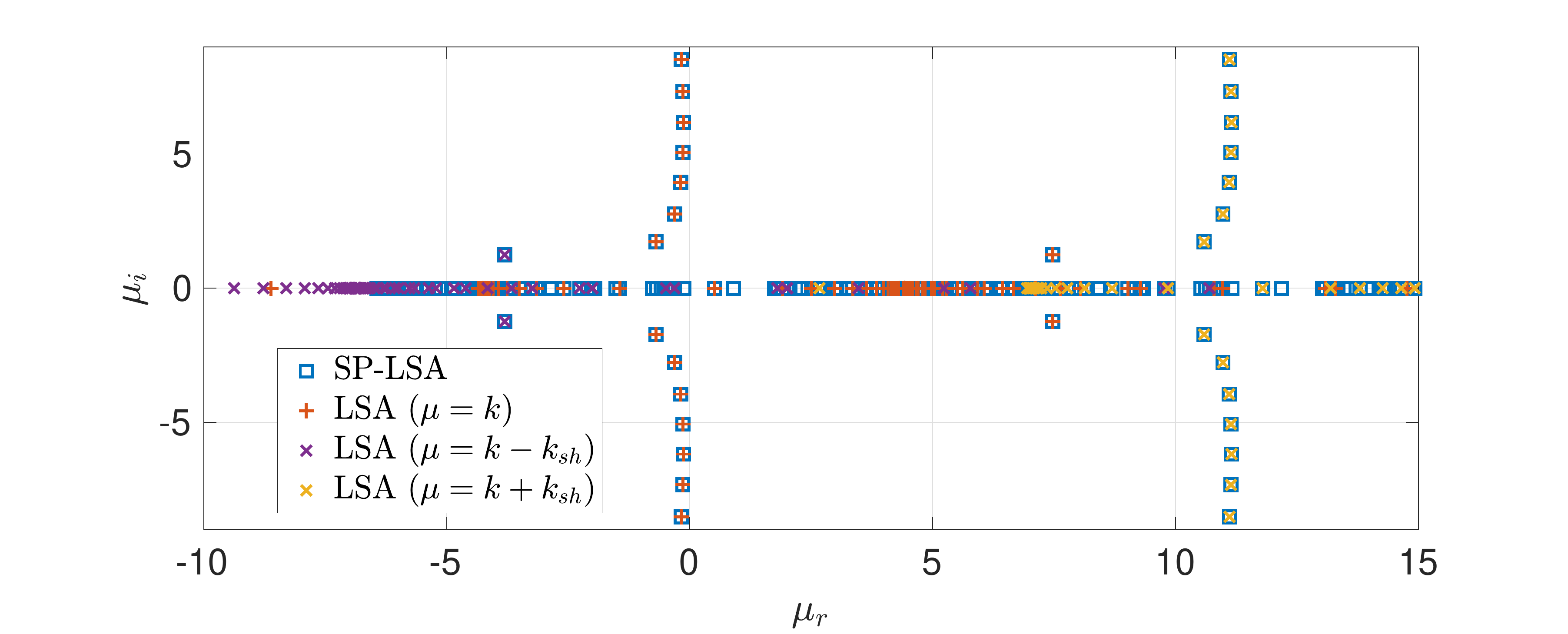}
\caption{Eigenspectrum containing 400 converged eigenvalues for $NPR=2.1$, $A_{sh}=0$, $St = 0.7$ (blue $\square$). The eigenspectrum of the locally parallel analysis is also shown for $\mu=k$ (orange $+$), $\mu=k-k_{sh}$ (purple $\times$), and $\mu=k+k_{sh}$ (yellow $\times$).}
\label{fig:EigsNPR21Ash0}
\end{figure}

As a first test, the equivalence of the $A_{sh}=0$ case with the locally parallel linear stability analysis (LSA) is demonstrated. To this end, the eigenspectrum for the $NPR=2.1$, $m=0$ case and $A_{sh}=0$ was compared to results from LSA using the same mean flow and $St=0.7$. The comparison between the eigenvalues is shown in figure \ref{fig:EigsNPR21Ash0}, where only 400 eigenvalues are shown. As expected, eigenvalues in the spectrum occur with a $k_{sh}$ periodicity, such that $\mu=\mu+N k_{sh}$, with $N$ an integer; this is observed more clearly in the modes close to the imaginary axis (the acoustic branch), that also appear close to $\mu_r=k_{sh}$, and in the unstable Kelvin-Helmholtz mode, now also present in the $\mu_r<0$ part of the spectrum due to this periodicity. Figure \ref{fig:EigsNPR21Ash0} also shows that the eigenvalues of the locally parallel stability align perfectly with those from the periodic case, considering the periodicity of the solution. Most modes could be captured by the LSA using $\mu=k$ and $\mu=k\pm k_{sh}$; the few modes without equivalence seen on the real axis are related to soft-duct modes of higher wavenumber \citep{towne_cavalieri_jordan_colonius_schmidt_jaunet_bres_2017}. Due to the periodicity of the solution, all modes from LSA are now observed inside the interval $0\leq \mu \leq k_{sh}$. Consequently, the identification of discrete modes and mode branches can be performed directly by continuation from the locally parallel case.

The introduction of a small value of $A_{sh}=0.02$ does not lead to significant changes in the eigenvalues for the values of $NPR$ analysed herein (variations in growth rate and wavenumber are less than $0.01\%$ for real-valued frequencies). However, the eigenfunctions are non-trivially modified by the introduction of the shock-cell structure. Here, we analyse this effect on two of the most dynamically significant waves in this flow. The first is the Kelvin-Helmholtz (KH) mode \citep{michalke1965spatially}, which is responsible for the appearance of large scale vortical structures in the flow called wavepackets \citep{jordan2013wave}, one of the key structures responsible for sound radiation in turbulent jets \citep{cavalieri2019amr}. The second is the upstream-travelling guided jet mode, first identified by \cite{tam_hu_1989}. Recent works \citep{gojon2018aiaa,edgington-mitchell_jaunet_jordan_towne_soria_honnery_2018,mancinelli2019} have shown that screech tones are observed within the frequency bands of existence of these waves, at least for the axisymmetric mode, suggesting that this wave might be responsible for closing the resonance loop in screeching jets. The presence of the guided jet mode was also identified experimentally by \cite{edgington-mitchell_jaunet_jordan_towne_soria_honnery_2018,EdgingtonMitchell2020}, which also highlights the importance of such waves in the dynamics of the jet. The identification of these waves in the spatially periodic framework is less straightforward than in the locally parallel case, especially for the guided jet mode. Due to the $k_{sh}$ periodicity of the spectrum, this mode will now be mixed with critical layer modes and soft-duct modes \citep{towne_cavalieri_jordan_colonius_schmidt_jaunet_bres_2017} on the $\mu_r$ axis. Since the eigenvalues for such small $A_{sh}$ do not change relative to the $A_{sh}=0$ case, an auxiliary run of the locally parallel case was used to identify the modes related to each wave. 

\begin{figure}
\centering
\includegraphics[clip=true, trim= 50 20 20 10, width=\textwidth]{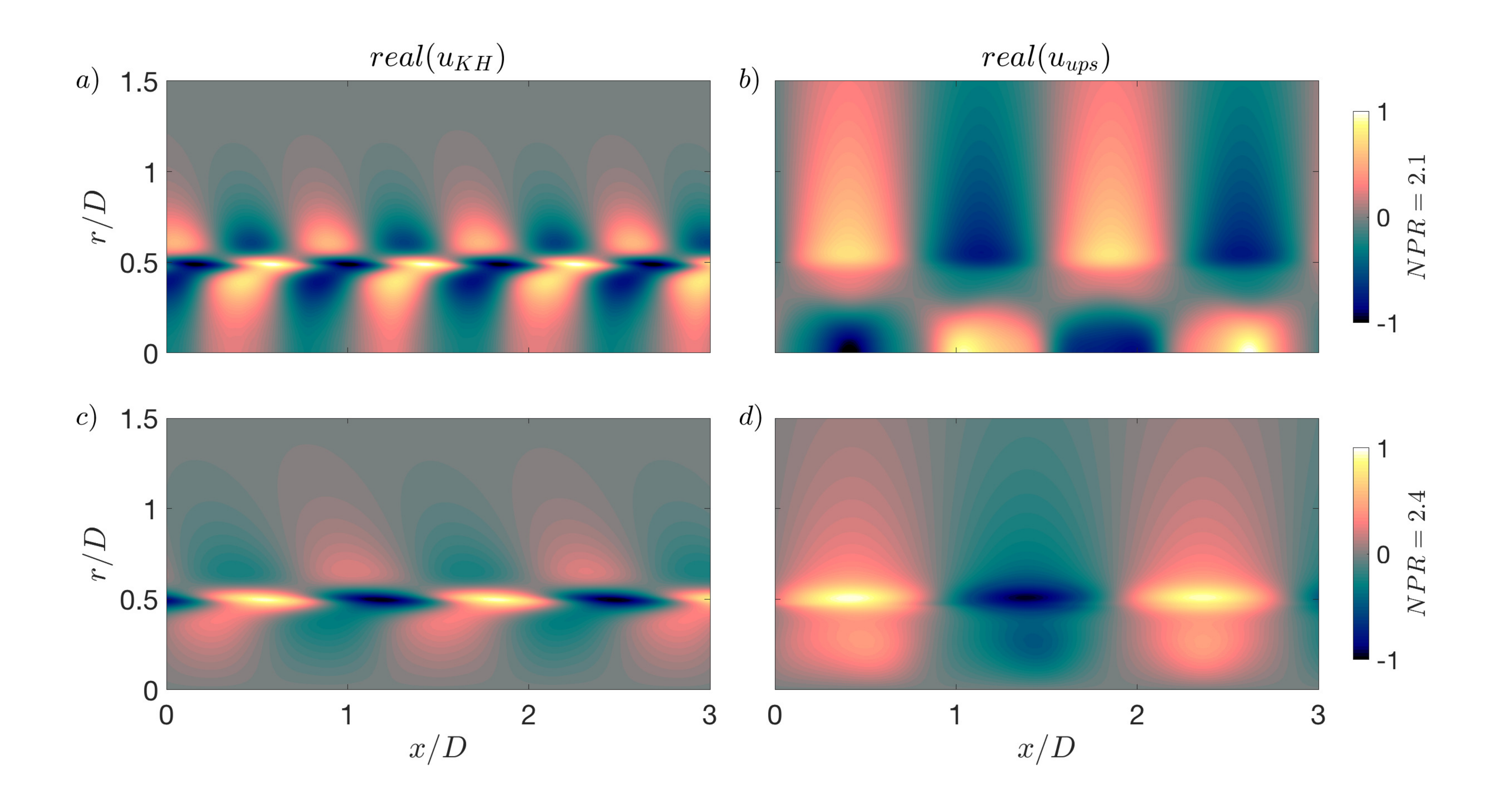}
\caption{Real part of the streamwise velocity of the different waves analysed. Top: Kelvin-Helmholtz (a) and upstream (b) modes for $m=0$, $NPR=2.1$ and $St=0.7$. Bottom: Kelvin-Helmholtz (c) and upstream (d) modes for $m=1$, $NPR=2.4$ and $St=0.47$. Spatially-periodic linear stability results for $A_{sh}=0.02$. The growth rates of the modes were removed to allow for a better visualisation of the radial structure.}
\label{fig:realUmodxr}
\end{figure}

The real part of the streamwise velocity of both the Kelvin-Helmholtz and upstream waves for $m=0,1$ and $St=0.7,0.47$  are shown in figure \ref{fig:realUmodxr}. The frequencies were chosen to be within the frequency range of existence of the neutral guided jet mode of radial order 2 and 1, respectively, and close to the screech frequencies observed in these cases (see \cite{edgington-mitchell_jaunet_jordan_towne_soria_honnery_2018,li_zhang_hao_he_2020}). Here, the modes are reconstructed spatially using (\ref{eqn:response_per}), and the imaginary part of the Floquet exponent was ignored to better visualise the oscillatory behaviour of the modes. The shapes of the Kelvin-Helmholtz modes follow the usual behaviour for $m=0$ and $m=1$ disturbances: in both cases, this wave is quite concentrated around the shear layer of the jet (since the instability is driven by shear), and a phase jump around this position is also observed. The differences between the $m=0$ and $m=1$ modes are observed mainly around the centreline, where the helical modes should reach zero amplitude due to their natural boundary conditions \citep{batchelor1962analysis}. Both Kelvin-Helmholtz modes depicted in figures \ref{fig:realUmodxr}(a,c) are comparable with the structures educed from a proper orthogonal decomposition (POD) of experimental data, previously presented in \cite{edgington-mitchell_jaunet_jordan_towne_soria_honnery_2018} ($m=0$ dominated) and \cite{edgington-mitchell_oberleithner_honnery_soria_2014} ($m=1$ dominated). The guided jet modes for the two cases are also shown in figure \ref{fig:realUmodxr} (b,d). These modes also follow the expected symmetry, being equivalent to the ones presented by \cite{gojon2018aiaa}, for each azimuthal wavenumber and radial order.  


\begin{figure}
\centering
\includegraphics[clip=true, trim= 50 20 20 10, width=\textwidth]{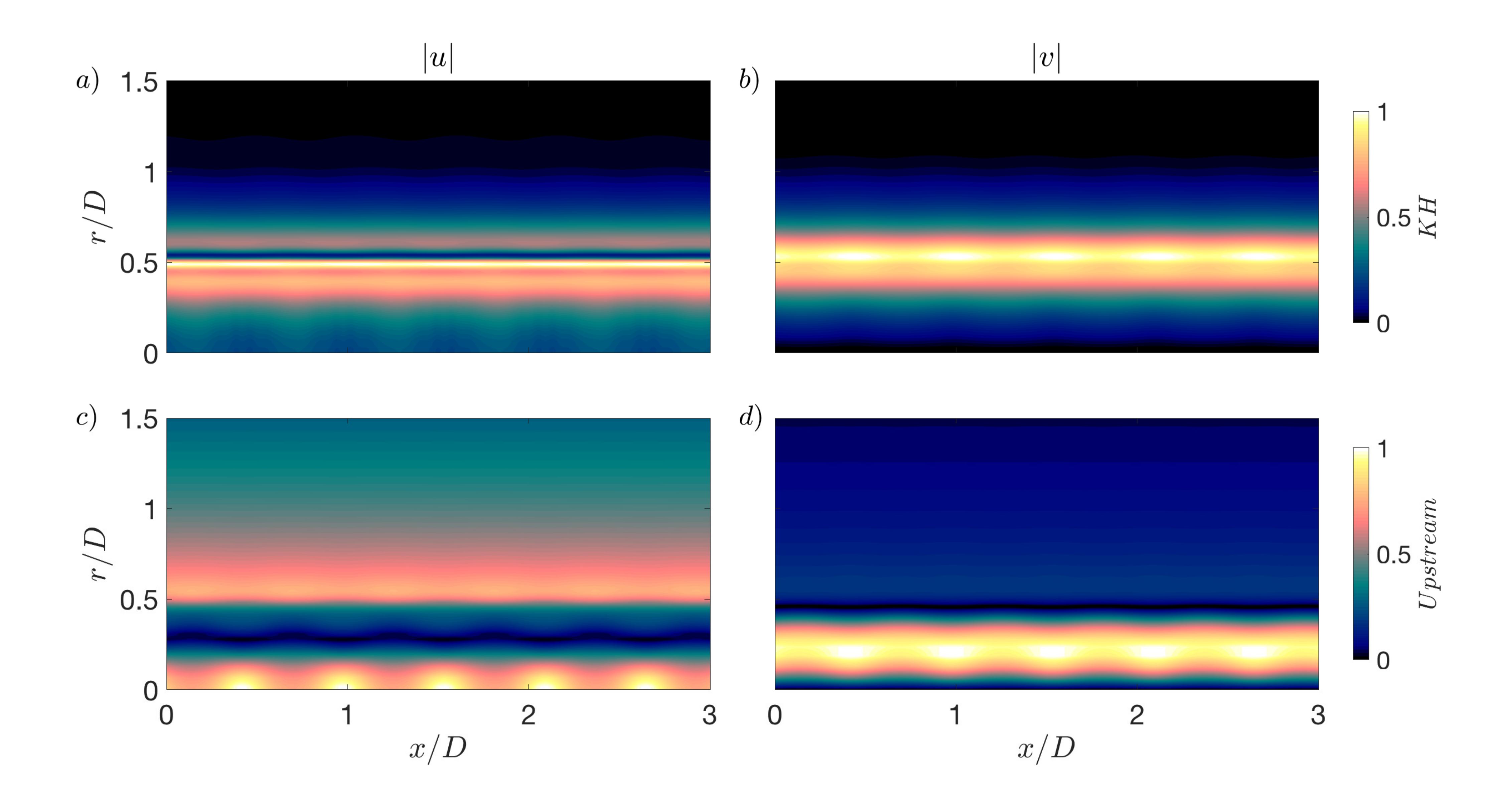}
\caption{Absolute value of the streamwise (a,c) and radial (b,d) velocities of the different waves analysed for $m=0$, $NPR=2.1$ and $St=0.7$. Kelvin-Helmholtz (a,b) and upstream (c,d) modes for $A_{sh}=0.02$. The growth rates of the modes were removed to allow for a better visualisation of the radial structure.}
\label{fig:absUmodxrm0}
\end{figure}

\begin{figure}
\centering
\includegraphics[clip=true, trim= 50 20 20 10, width=\textwidth]{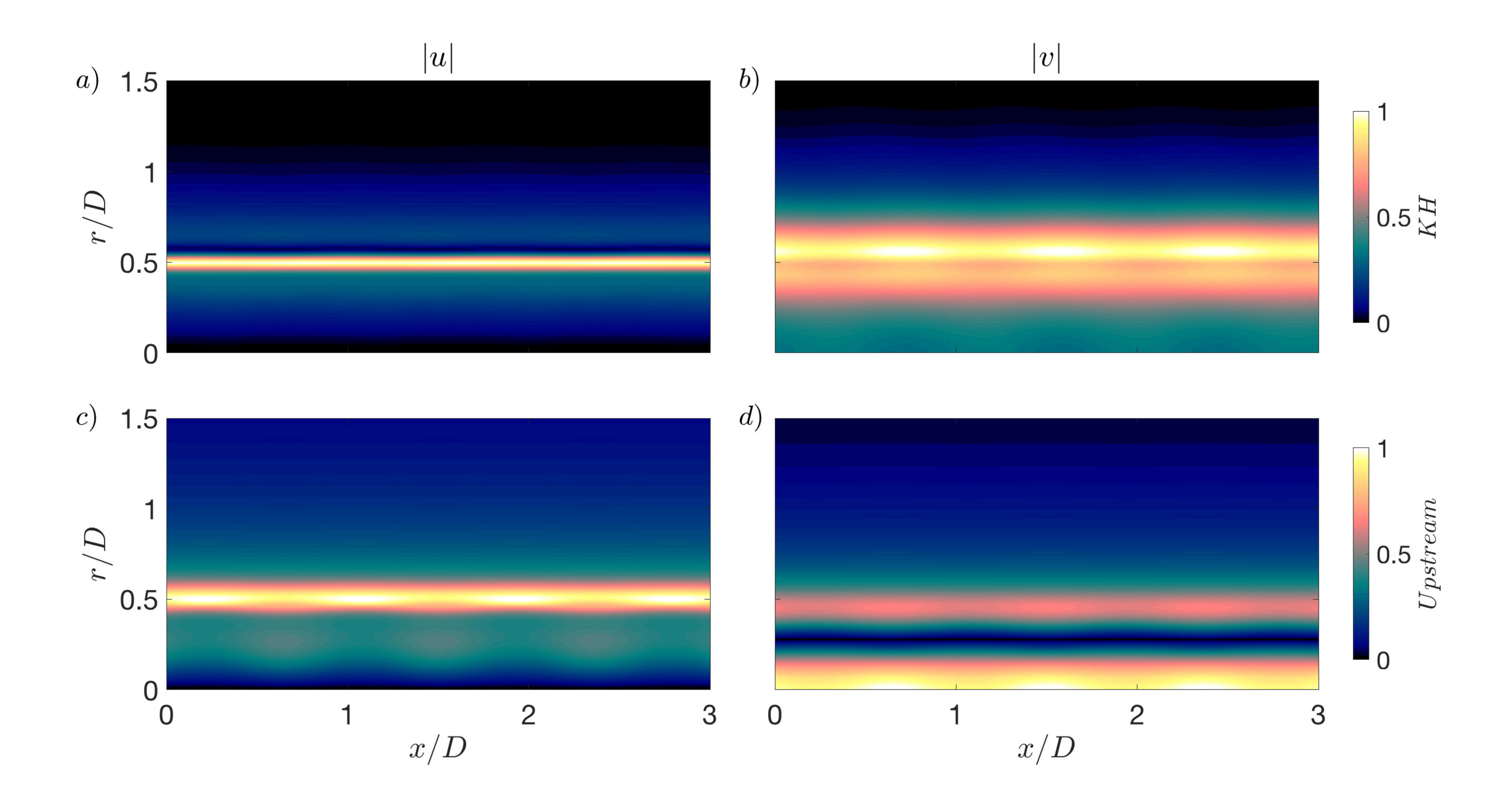}
\caption{Absolute value of the streamwise (a,c) and radial (b,d) velocities of the different waves analysed for $m=1$, $NPR=2.4$ and $St=0.47$. Kelvin-Helmholtz (a,b) and upstream (c,d) modes for $A_{sh}=0.02$. The growth rates of the modes were removed to allow for a better visualisation of the radial structure.}
\label{fig:absUmodxrm1}
\end{figure}

Figure \ref{fig:realUmodxr} yields little insight regarding the effect of the shock-cell structure on the different modes, since the modulation is masked by the phase evolution of the waves. This effect can be seen more clearly by looking at the absolute value of the velocity for each case. The locally parallel stability results lead to modes without any streamwise variation in the absolute value of all flow quantities, while the spatially periodic case allows the variation following the flow periodicity. This is shown in figures \ref{fig:absUmodxrm0} ($m=0$) and \ref{fig:absUmodxrm1} ($m=1$) for the streamwise and radial velocity components. The changes in the streamwise velocity of the axisymmetric modes induced by the inclusion of the periodic shock-cell structure are mainly concentrated around the centreline; the periodic shocks have little effect around the shear layer. This changes when the radial velocity of the Kelvin-Helmholtz is considered: modulation for this quantity actually occurs around the shear layer. The modulation of the radial velocity for the upstream mode occurs between the centreline and the first node in the radial direction (where the phase shift occurs). 

The behaviour of the modulated helical disturbances is shown in figure \ref{fig:absUmodxrm1}. For this case, a slight modulation is observed in the streamwise velocity of the Kelvin-Helmholtz mode, mostly around the shear layer and in the outer region of the jet. The modulation is stronger in the radial velocity of the KH wave and in both components of the upstream wave. Interestingly, different regions of the jet respond differently to the presence of the periodic structure; while some regions of the jet follow the same oscillatory behaviour as the shock-cell structure, other regions may follow the periodicity with a phase shift. This is observed, for example, in figure \ref{fig:absUmodxrm1}(c), where disturbances around the shear layer follow an inverse behaviour compared to disturbances at the centreline. Also, for both $m=0$ and $m=1$ disturbances, the modulation of the upstream-travelling waves is stronger than for the Kelvin-Helmholtz mode.

\begin{figure}
\centering
\includegraphics[clip=true, trim= 50 0 100 0, width=\textwidth]{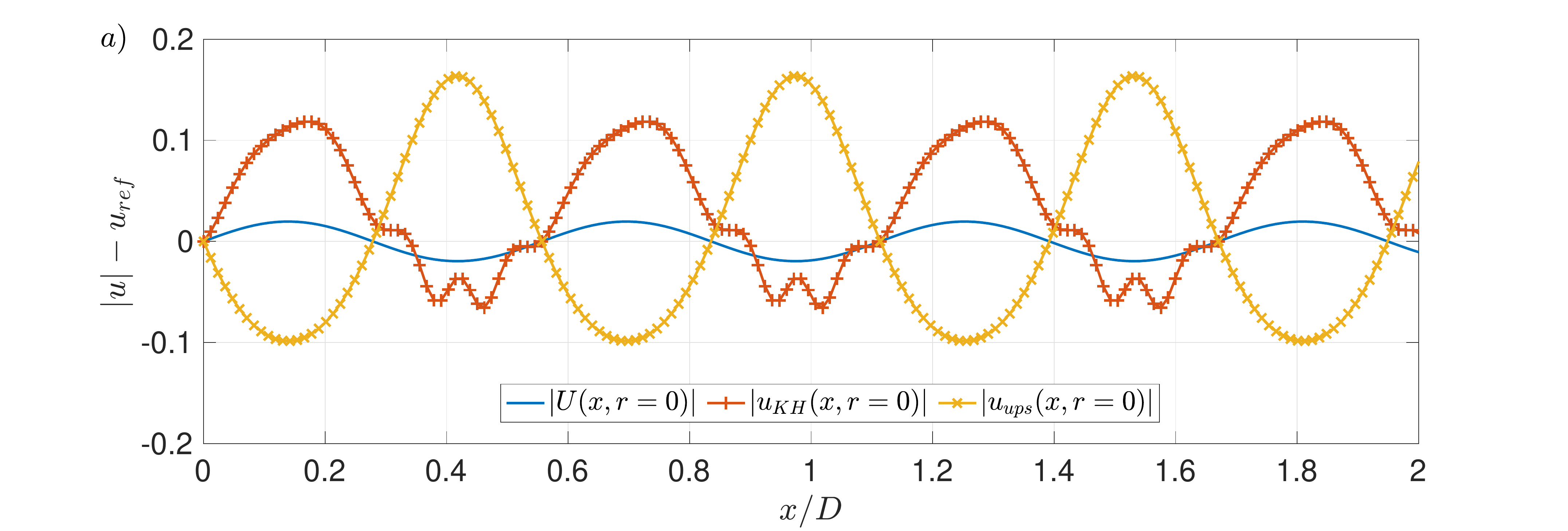}
\includegraphics[clip=true, trim= 30 0 110 0, width=0.93\textwidth]{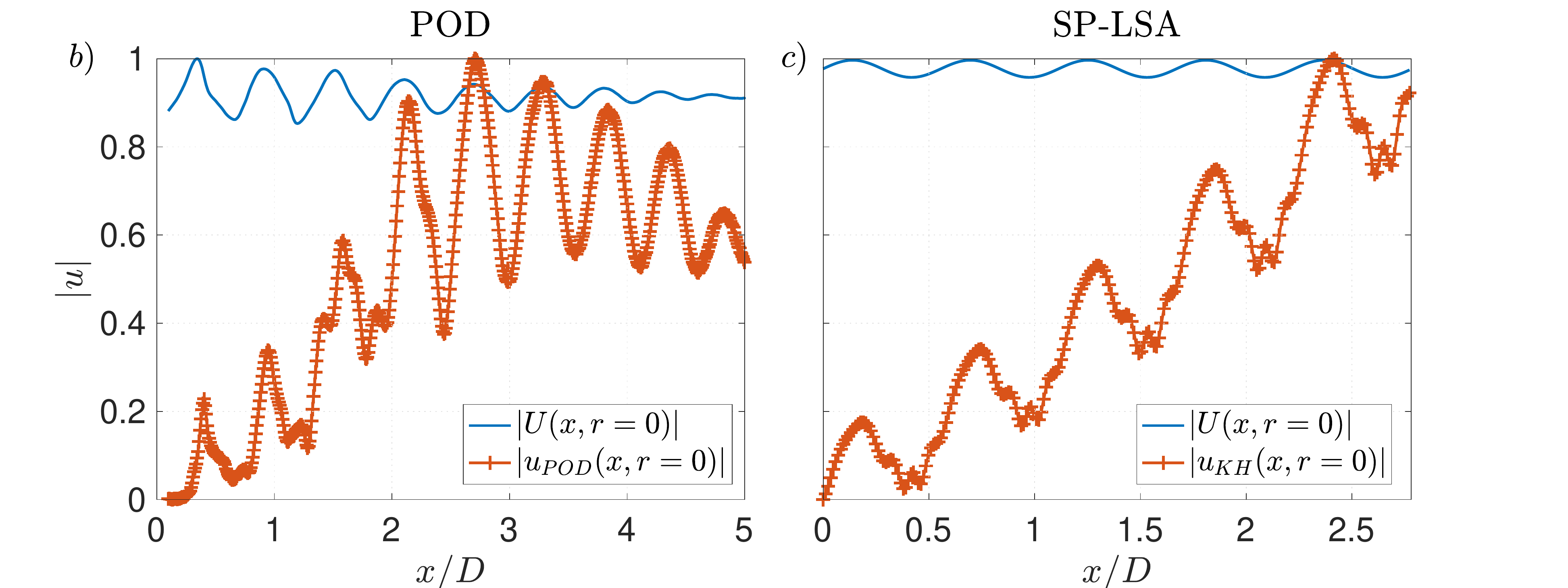}
\caption{Comparison between the modulation of the different waves in the spatially periodic linear stability analysis and the POD mode from \cite{edgington-mitchell_jaunet_jordan_towne_soria_honnery_2018} at the centreline ($r/D=0$) and $NPR=2.1$. The spatially periodic modes (a) are normalised by their maximum and subtracted from their value at $x/D=0$, and the imaginary part of the Floquet exponent was ignored in the spatial reconstruction of the modes. In (b), the POD mode and mean streamwise velocity are normalised by their maximum. The mean velocity and the spatially growing KH mode (growth rate reduced to $20\%$ of its original value) (c) are also normalised by their maximum.}
\label{fig:Umodcentre21}
\end{figure}

Figure \ref{fig:Umodcentre21}(a) shows the modulation in streamwise velocity of both waves studied herein for $NPR=2.1$, $m=0$, $St=0.7$ and $r/D=0$; the mean velocity at the centreline is also shown for reference. All fields were normalised by their maximum and subtracted from the initial value at $x/D=0$. This plot shows clearly that the shock-cell structure has opposing effects on the modulation of the different waves in the flow: while a maximum in the mean velocity is associated with a maximum in the streamwise velocity of the Kelvin-Helmholtz wave, it is also associated with a minimum of the guided jet mode. Considering that these waves travel in opposite directions, this difference is possibly associated with the effect of a wave passing through a shock coming from regions downstream or upstream of it. Figure \ref{fig:Umodcentre21}(a) confirms that the modulation effect is stronger in the upstream-travelling wave for the $m=0$ case. While the modulation of the guided jet mode is rather simple in shape, the modulation of the Kelvin-Helmholtz mode is more intricate. A slight oscillation is observed around the mean point between a maximum and a minimum (or a node) of the mean velocity, and a much stronger oscillation occurs at the position of minimum velocity.

A direct comparison between the present results and experiments is hard to perform. Usually, the lack of time-resolved data only allows the identification of the structure related to the resonance process (the screech mode) using proper orthogonal decomposition or other similar modal decomposition tools \citep{edgington-mitchell_oberleithner_honnery_soria_2014,edgington-mitchell_jaunet_jordan_towne_soria_honnery_2018,li_zhang_hao_he_2020}. However, the screech mode is composed of several downstream- and upstream-travelling waves, which hinders the evaluation of such modulation effects in the different waves of the flow. Still, considering that the screech mode is dominated by a Kelvin-Helmholtz wavepacket, it could be expected that some of the trends identified in the present study should be present in POD modes of experimental data. In order to evaluate this, a complex mode is built using the two first POD modes, as in \cite{EdgingtonMitchell2020} for $NPR=2.1$; the details about the experiments and numerical procedure can be found in the cited paper. The streamwise evolution of the screech mode at the centreline is shown in figure \ref{fig:Umodcentre21}(b), where the mean velocity is also shown as reference. Comparing figures \ref{fig:Umodcentre21}(a) and (b), some similar features can be identified. As in the spatially-periodic analysis, the POD mode also has maxima in streamwise velocity aligned with maxima in the mean velocity (maxima of the shock-cell structure). Also, oscillations following the same pattern are observed around the third shock-cell, both associated with nodes and minima in the mean velocity. Considering a spatial growth of the KH mode enhances the comparison, as shown in figure \ref{fig:Umodcentre21}(c). However, there is no reason to expect quantitative agreement in growth rate between an inviscid, linear, periodic model and the experimental data; indeed, there is a significant difference between the values. As the purpose of this figure is purely to illustrate similar trends in the modulation effect, the growth rate of the KH mode is reduced to $20\%$ of its value for the purpose of this comparison. Overall, considering that the only spatial evolution allowed in the present model is the periodic shock-cell structure, the present model captures some of the experimental trends quite closely, supporting the use of the present tool in the analysis of shock-containing flows.

\subsection{Saddle points and absolute instability}

In the previous section, we showed that the spatially periodic linear stability analysis manages to capture some of the trends identified in experiments. We also showed that results from this formulation can be analysed in the same fashion as the locally parallel case, with modes categorised in a very similar manner to this previous method; in fact, for $A_{sh}=0$, the present method recovers the exact same modes as the local analysis. As mentioned in the previous section, small changes in the growth rates and peak wavenumber were identified for $A_{sh}=0.02$. Thus, one could erroneously think that the only effect of such flow periodicity is on the shapes of the different waves supported by the flow. In this section we show that this is not the case. In fact, the periodicity changes the nature of the instability mechanism to which the flow is subject, and this will have a direct effect on the physical interpretation of the screech phenomenon. 

In the present section, we follow the formulation developed by \cite{huerre_monkewitz_1985}, revised in \cite{huerre1990local}, and extended to spatially periodic base flows in \cite{Brevdo1996}. For a clearer interpretation of the results of \cite{Brevdo1996}, the reader can also refer to \cite{BrancherChomaz1997}. In summary, an absolute instability occurs if a saddle in the $(\omega,\mu)$ plane exists for $\omega_{0,i}>0$, and this saddle must be formed between an upstream- and a downstream-travelling mode. Considering that all waves are now in the same region of the spectrum (between $0\leq \mu_r \leq k_{sh}$), an interaction between downstream- and upstream-travelling waves is more likely to occur. In order to evaluate this, the complex frequencies $\omega$ in which the complex eigenvalue related to the Kelvin-Helmholtz mode approaches the eigenvalue of an upstream travelling wave were sought for the locally parallel case. Then, second and third runs of the spatially periodic stability were performed for $A_{sh}=0$ and $A_{sh}=0.02$, in order to evaluate the effect of the mean flow periodicity, checking if the shock-cell structure leads to an interaction between the different modes. In the parametric studies, saddles were sought using the methodology proposed by \cite{monkewitz_1988}.

\begin{figure}
\centering
\includegraphics[clip=true, trim= 0 0 0 0, width=\textwidth]{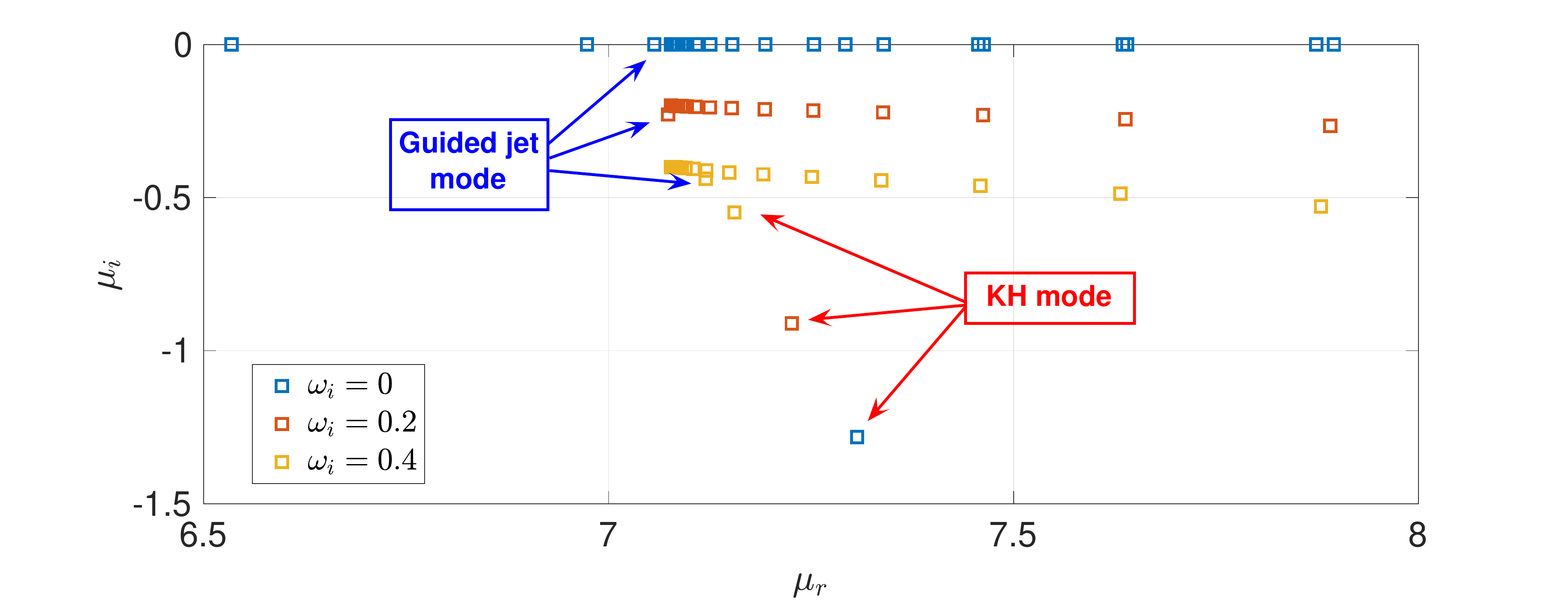}
\caption{Trajectories of the different modes in the complex $\mu$ plane, for $NPR=2.1$, $m=0$, $A_{sh}=0.02$ and $St=0.685$, for several values of $\omega_i$.}
\label{fig:OmegaiEffect}
\end{figure}

The effect of increasing $\omega_i$ in the eigenspectrum is exemplified in figure \ref{fig:OmegaiEffect}. The Kelvin-Helmholtz mode is directly identified as the only unstable mode ($\mu_i<0$) for $\omega_i=0$. As expected for an unstable downstream-propagating mode, increasing $\omega_i$ leads to an increase in the value of $\mu_i$; for larger imaginary frequencies, this mode will cross the real axis and remain in the $\mu_i>0$ region of the spectrum. The opposite happens with the guided jet mode, identified as the discrete mode closest to the continuous neutral acoustic branch for $\omega_i=0$. Since this mode is upstream propagating, increasing the imaginary frequency leads to a decrease in $\mu_i$. The contrasting behaviour of these two modes leads them to be located quite close to each other for some values of $\omega_i$, which could allow for an interaction between the downstream- and upstream-travelling modes.

\begin{figure}
\centering

\includegraphics[clip=true, trim= 0 0 0 0, width=0.5\textwidth]{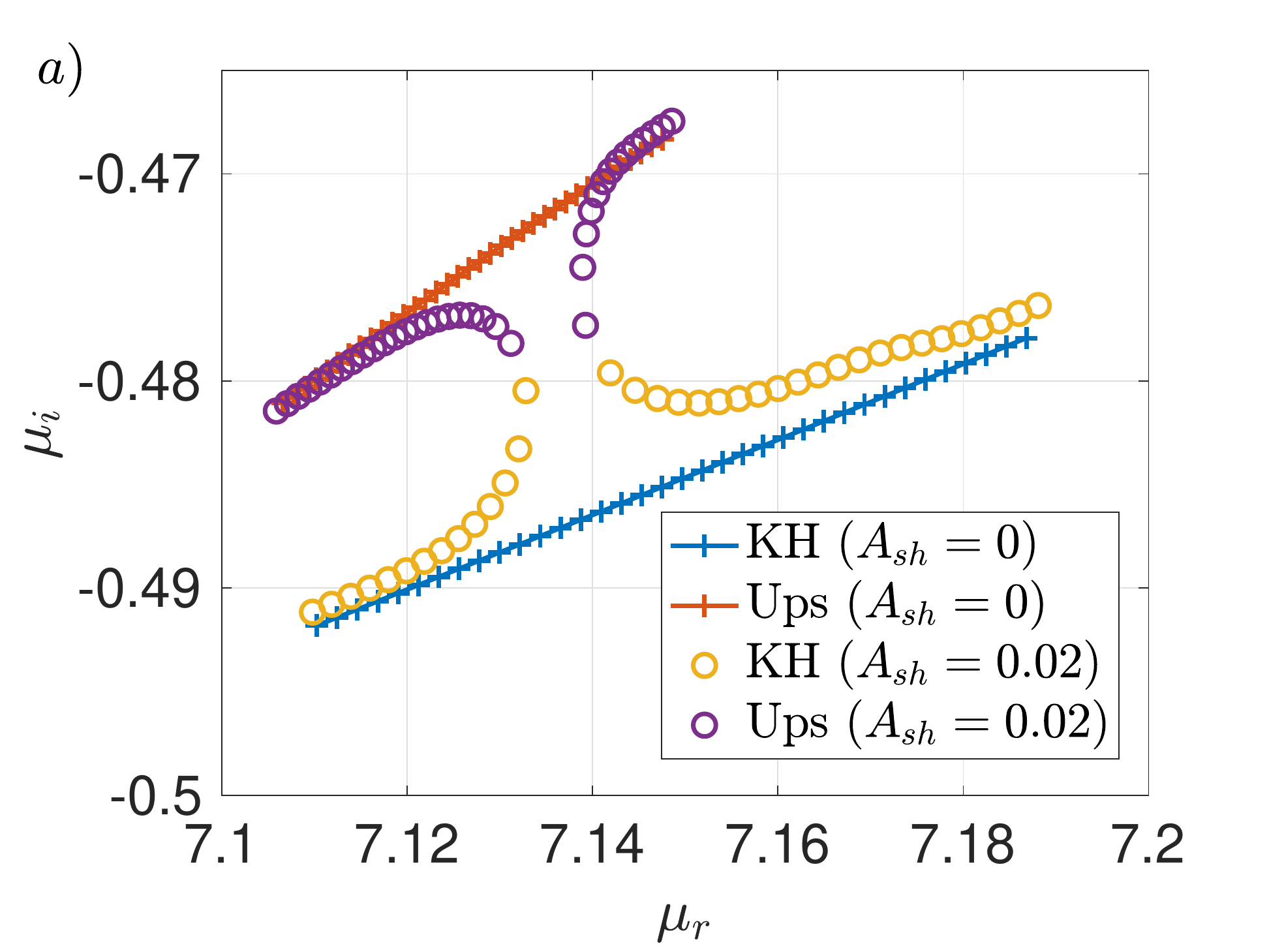}\includegraphics[clip=true, trim= 0 0 0 0, width=0.5\textwidth]{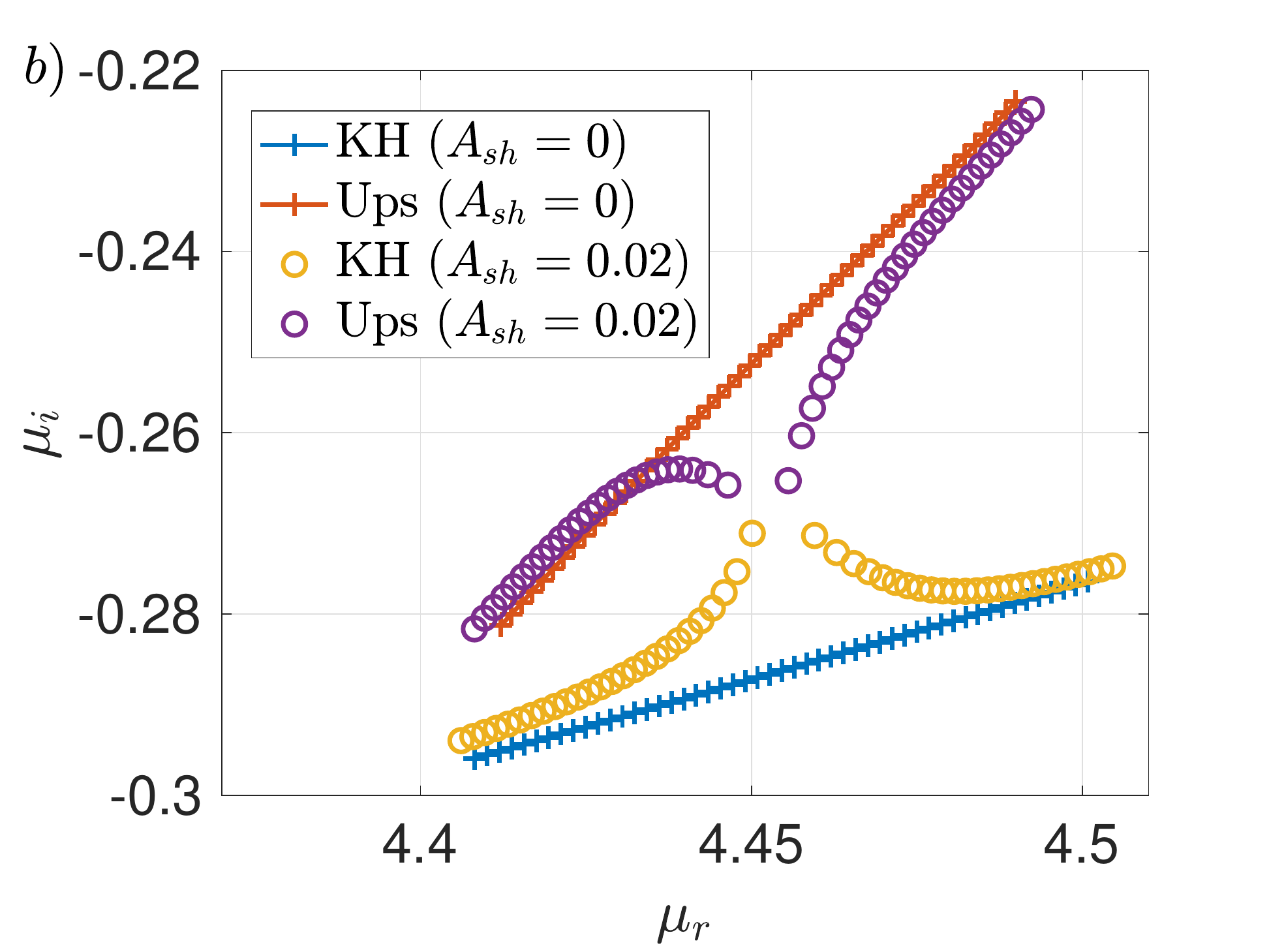}
\includegraphics[clip=true, trim= 0 0 0 0, width=0.5\textwidth]{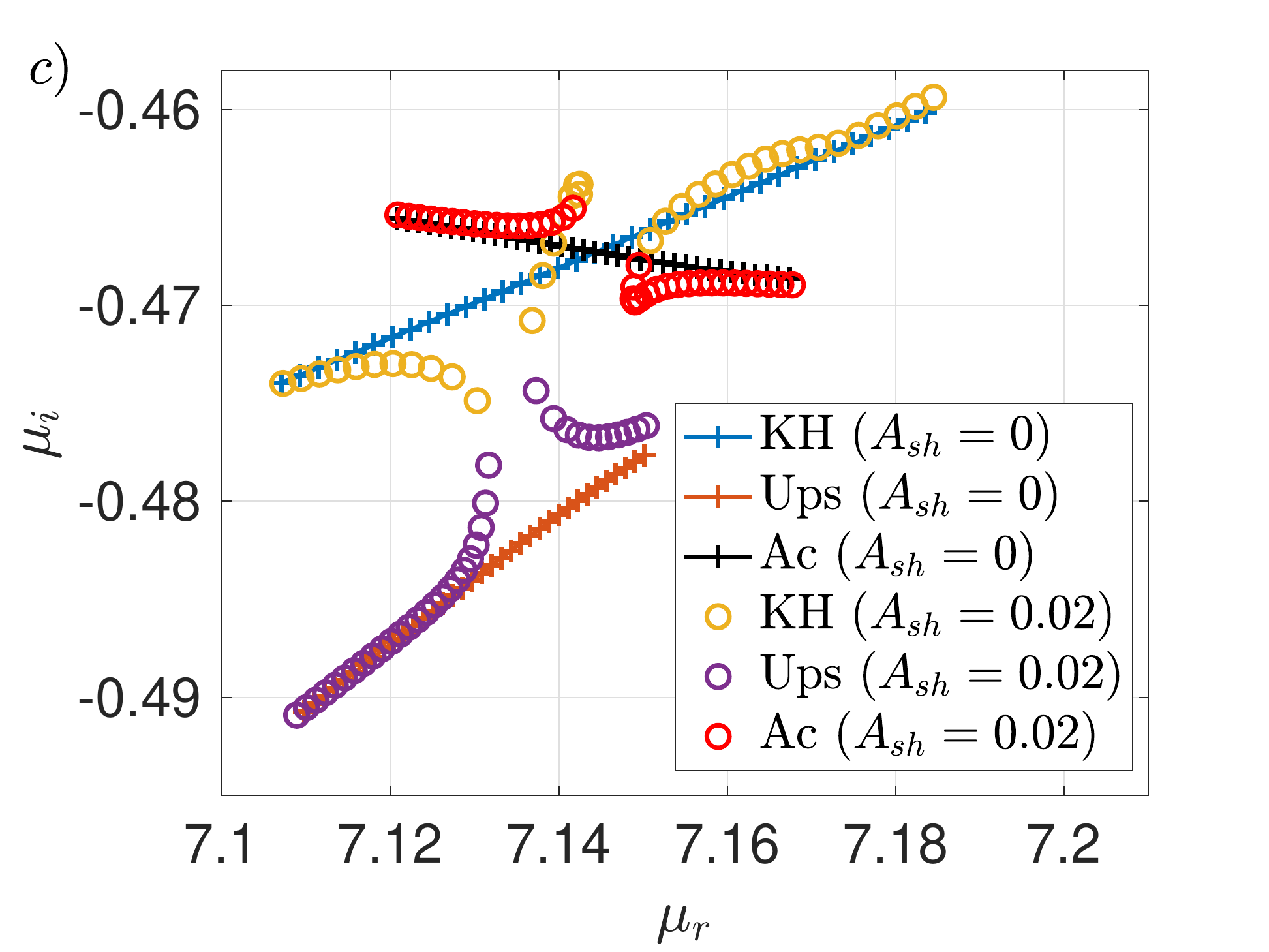}\includegraphics[clip=true, trim= 0 0 0 0, width=0.5\textwidth]{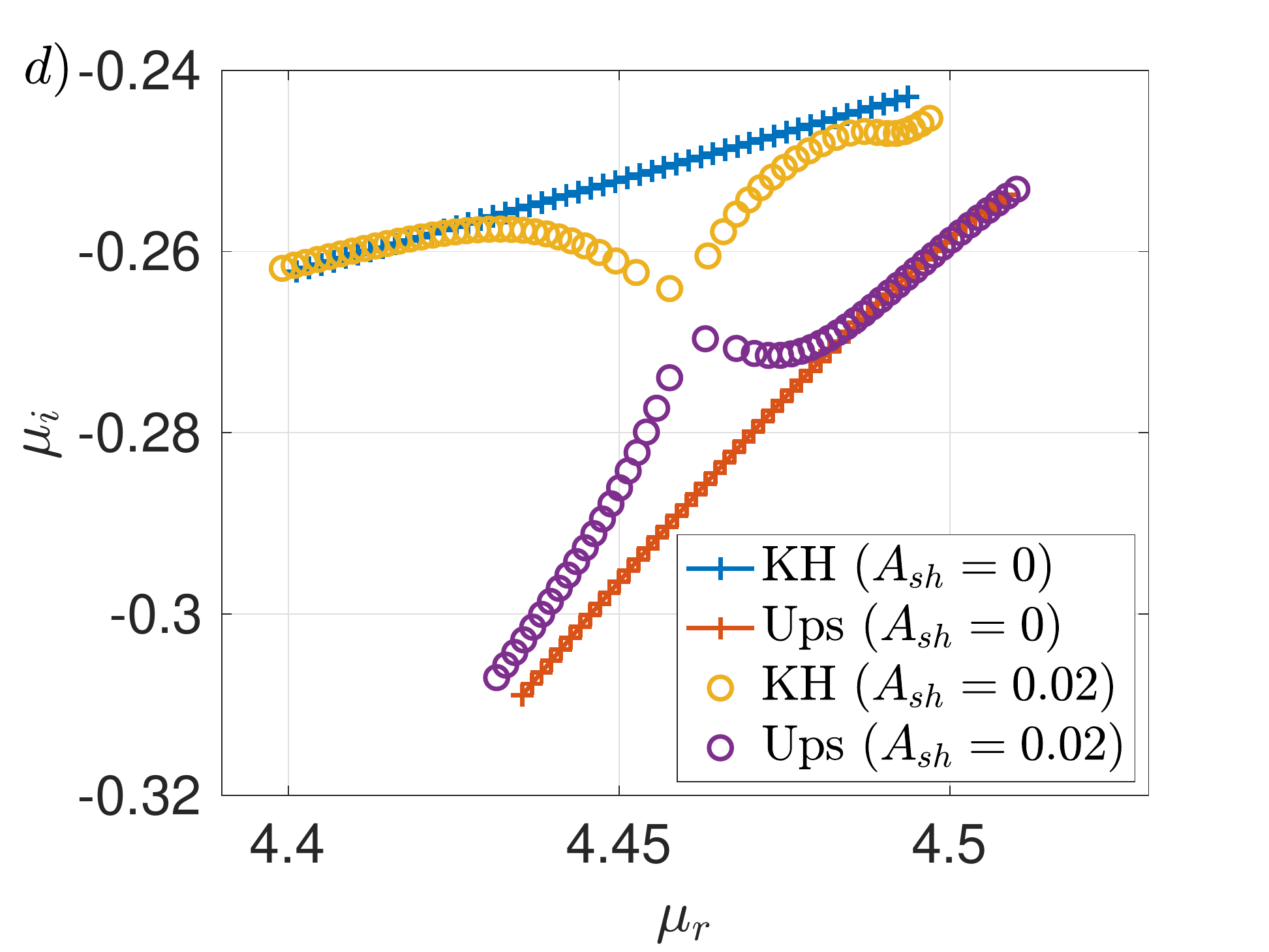}
\caption{Trajectories of the Kelvin-Helmholtz (KH) and guided jet (Ups) modes in the complex $\mu$ plane, for both $m=0$ (a,c) and $m=1$ (b,d) cases, for $\omega_i=0.436$ (a), $\omega_i=0.180$ (b), $\omega_i=0.446$ (c) and $\omega_i=0.204$ (d). Modes are shown in the interval $0.682 \leq St \leq 0.689$ for the axisymmetric case, and $0.442 \leq St \leq 0.452$ for the helical case. All modes are named according to their identity at the lowest Strouhal number in each plot. In (c), the trajectory of a mode from the continuous acoustic spectrum (Ac) is also shown.}
\label{fig:SaddlesAll}
\end{figure}

The trajectories of the eigenvalues related to Kelvin-Helmholtz (KH) and guided-jet/upstream (Ups) modes in the complex $\mu$ plane for increasing $St$ number are shown in figure \ref{fig:SaddlesAll}. These trajectories were computed for $\omega_i=\{ 0.436,0.446 \}$ ($m=0$) and $\omega_i=\{0.180, 0.204\}$ ($m=1$), and in the interval $0.682 \leq St \leq 0.689$ and $0.442 \leq St \leq 0.452$ respectively. In both figures \ref{fig:SaddlesAll}(a,b), KH modes travel from left to right for increasing $St$, and upstream modes move from right to left. For $A_{sh}=0$, the trajectories of both modes approach each other, but continue straight in their original path; considering that the $A_{sh}=0$ case is exactly equivalent to the locally parallel case (except for the periodicity of the modes in the complex $\mu$ plane), this is not surprising: no absolute instability is observed between these two modes, as they originally inhabited different regions of the spectrum in the locally parallel case. As the periodicity is included ($A_{sh}=0.02$), the trajectories of these modes deform around a given point $\mu_s$ of the spectrum. Close to this point, both modes move towards each following a specific direction, and are repelled in a perpendicular direction for higher $St$. This is a direct symptom of a double root in the $\mu(\omega)$ relation, characterising a saddle point. As in the locally parallel case, these two waves move in opposite directions of the real $\mu$-axis for increasingly large $\omega_i$, being equivalent to $k^+$ (KH mode) and $k^-$ (upstream mode) waves (see \cite{huerre1990local} for more details). In the spatially periodic framework, this also means that for $\omega_i \to \infty$, one of the modes travels internally to the unit circle for the Floquet multiplier ($|\ee^{\ii \mu \lambda_{sh}}|=1$) and the other travels externally to that circle. These two conditions, associated with the fact that the saddles occur for $\omega_{0,i}>0$, define the absolute instability of a spatially periodic system, as derived by \cite{Brevdo1996}. Thus, the impulse response of the spatially periodic shock-cell pattern grows in both downstream and upstream directions, leading to an oscillator behaviour that spreads throughout the domain for large times.

Considering that these modes have similar trajectories for several values of $\omega_i$, this phenomenon occurs twice for this periodic flow: once when the KH mode is on the bottom part of the spectrum (low $\omega_{0,i}$, figures \ref{fig:SaddlesAll}(a,b)), and again when this mode is at the top part of the spectrum (higher $\omega_{0,i}$, figures \ref{fig:SaddlesAll}(c,d)). These two saddles are formed by the same modes, and are very close to one another. The spatio-temporal growth rates associated with these are also similar, suggesting that both saddles are representative of the same phenomenon. With that in mind, mode shapes and wavenumber spectra displayed in the remainder of this paper are related to the second saddle (higher $\omega_{0,i}$), as it dominates the impulse response of the flow for large $t$.

An interaction between the KH mode and a mode from the acoustic continuous spectrum is also observed in figure \ref{fig:SaddlesAll}(c). Given that the upstream component of screech was historically assumed to be characterised by free stream acoustic waves, such interaction may at first seem of significance. However, care must be taken in its interpretation for a number of reasons. Modes from this branch are discrete representations of a continuous branch associated with sound waves radiating in several directions \citep{gloor2013linear}, and are related to a branch cut in the Briggs-Bers analysis, as shown by \cite{huerre_monkewitz_1985}. In this previous work, the authors have also shown that the contribution of this continuous branch decays exponentially for large times, and was neglected in the absolute instability analysis. Also, as the KH and guided jet modes switch identities at the saddle, it is hard to determine if the interaction between the discrete mode and the acoustic branch is formed between downstream- and upstream-travelling waves; thus, such interaction may be entirely due to the partial identification of this mode as a guided jet mode, which ejects from the acoustic branch for a given frequency (see \cite{towne_cavalieri_jordan_colonius_schmidt_jaunet_bres_2017}), and may still be able to interact with such branch. No saddle with acoustic modes was found for imaginary frequencies far from the saddles related to the guided jet mode.

Figures \ref{fig:SaddlesAll}(a-d) show that saddle-points between Kelvin-Helmholtz and guided jet modes can be found in spatially periodic flows, such as shock-containing jets; unlike for the ideally expanded case where an absolute instability is only observed in low-density jets \citep{monkewitz1988absolute}, the presence of a periodic structure, such as the train of shock-cells, can allow for the interaction of modes that could not interact in the locally parallel framework. The connection between the present absolute instability mechanism and screech can be accessed by the analysis of both the frequency of the saddle (or the frequency of interaction between the two modes), and the shapes of the modes close to the saddle, keeping in mind the approximations of the current model. Table \ref{tab:Stcomp} shows the comparison between the frequencies of the saddles, reported in terms of the Strouhal number $St_{0}=\omega_{0,r}/(2\pi M)$, for both $m=0$ and $m=1$ cases, and the experiments of \cite{edgington-mitchell_jaunet_jordan_towne_soria_honnery_2018} and simulations of \cite{li_zhang_hao_he_2020}. Interestingly, the frequencies predicted by the spatially periodic analysis are quite close to the ones from these previous works, for both values of $NPR$ and azimuthal wavenumbers. This agreement supports the hypothesis that the absolute instability mechanism captures some features of the screech phenomenon, and that the frequency of the resonance loop is actually given by the frequency where the saddle point is identified. 

\begin{table}
  \begin{center}
\def~{\hphantom{0}}
  \begin{tabular}{cccc}
        $NPR$  & Spatially periodic analysis & Experiments/Simulation & Reference  \\[3pt]
        $2.1$ & $St_{0} \approx 0.684$ & $St_{screech}=0.67$ & \cite{edgington-mitchell_jaunet_jordan_towne_soria_honnery_2018}  \\ 
        $2.4$ & $St_{0} \approx 0.447$ & $St_{screech}=0.4237$ & \cite{li_zhang_hao_he_2020}
  \end{tabular}
  \caption{Approximate frequencies of the saddle in the spatially periodic analysis compared with screech frequencies from simulations/experiments.}
  \label{tab:Stcomp}
  \end{center}
\end{table}

\begin{figure}
\centering
\includegraphics[clip=true, trim= 0 0 0 0, width=0.5\textwidth]{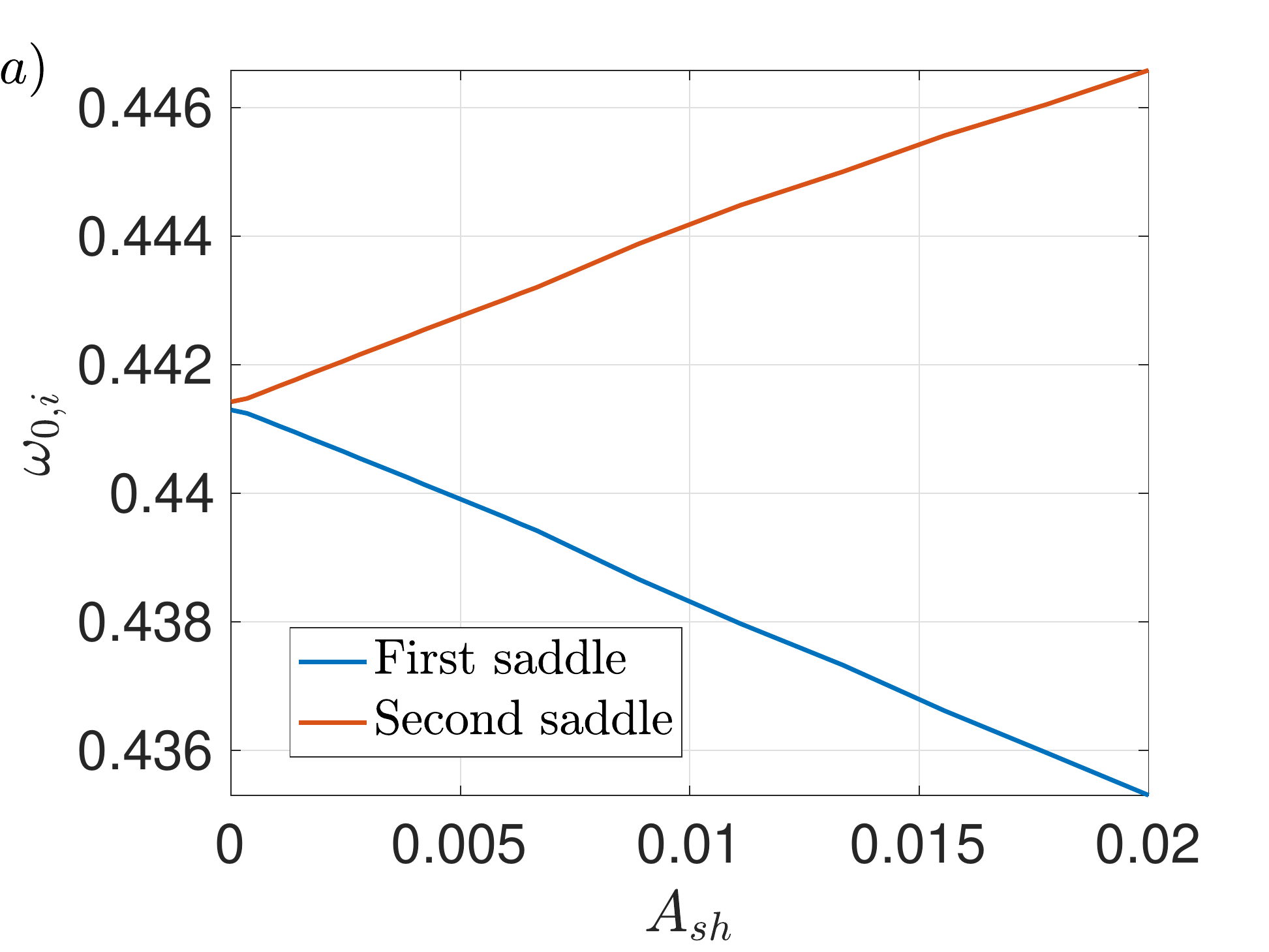}\includegraphics[clip=true, trim= 0 0 0 0, width=0.5\textwidth]{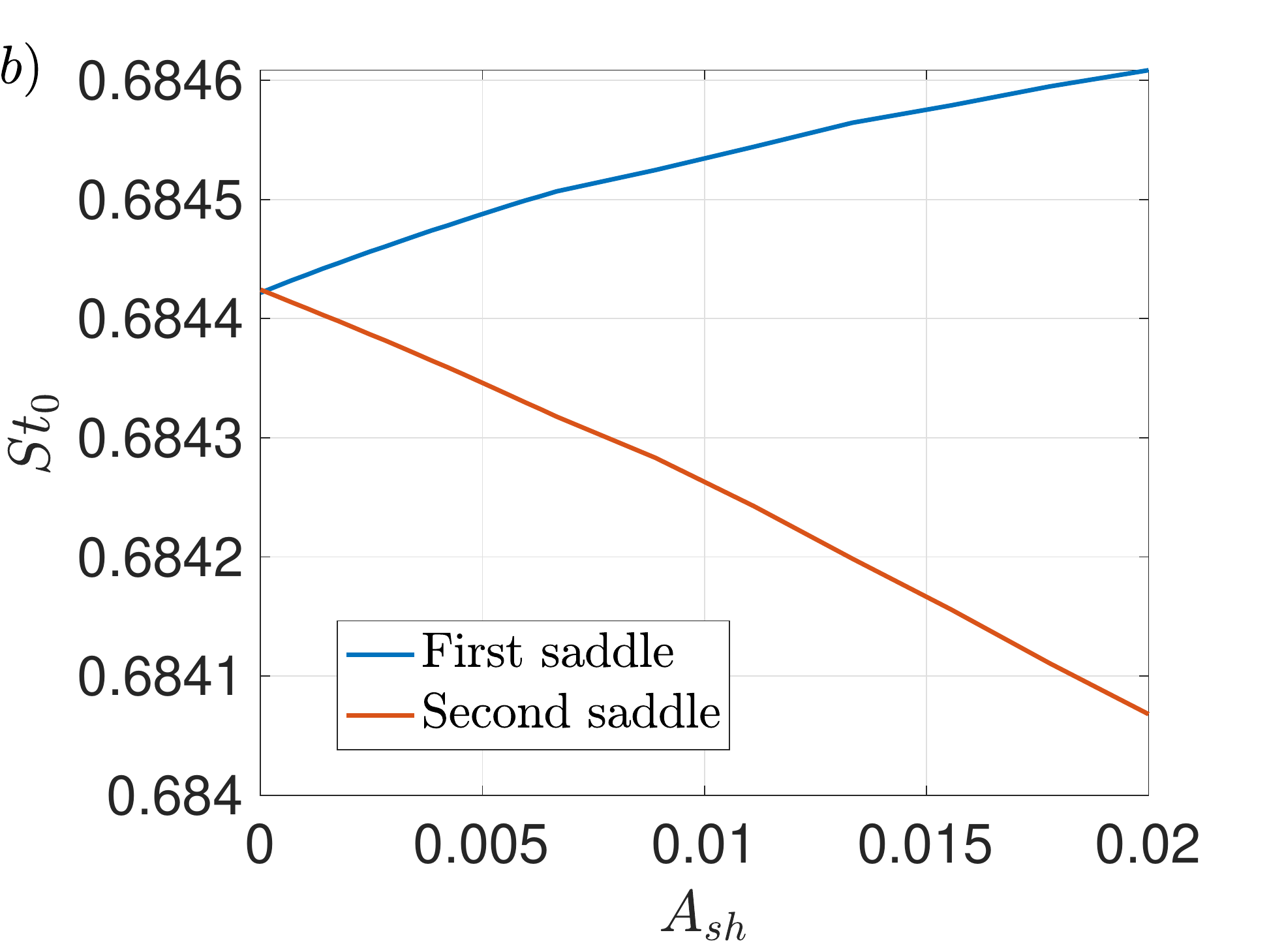}
\includegraphics[clip=true, trim= 0 0 0 0, width=0.5\textwidth]{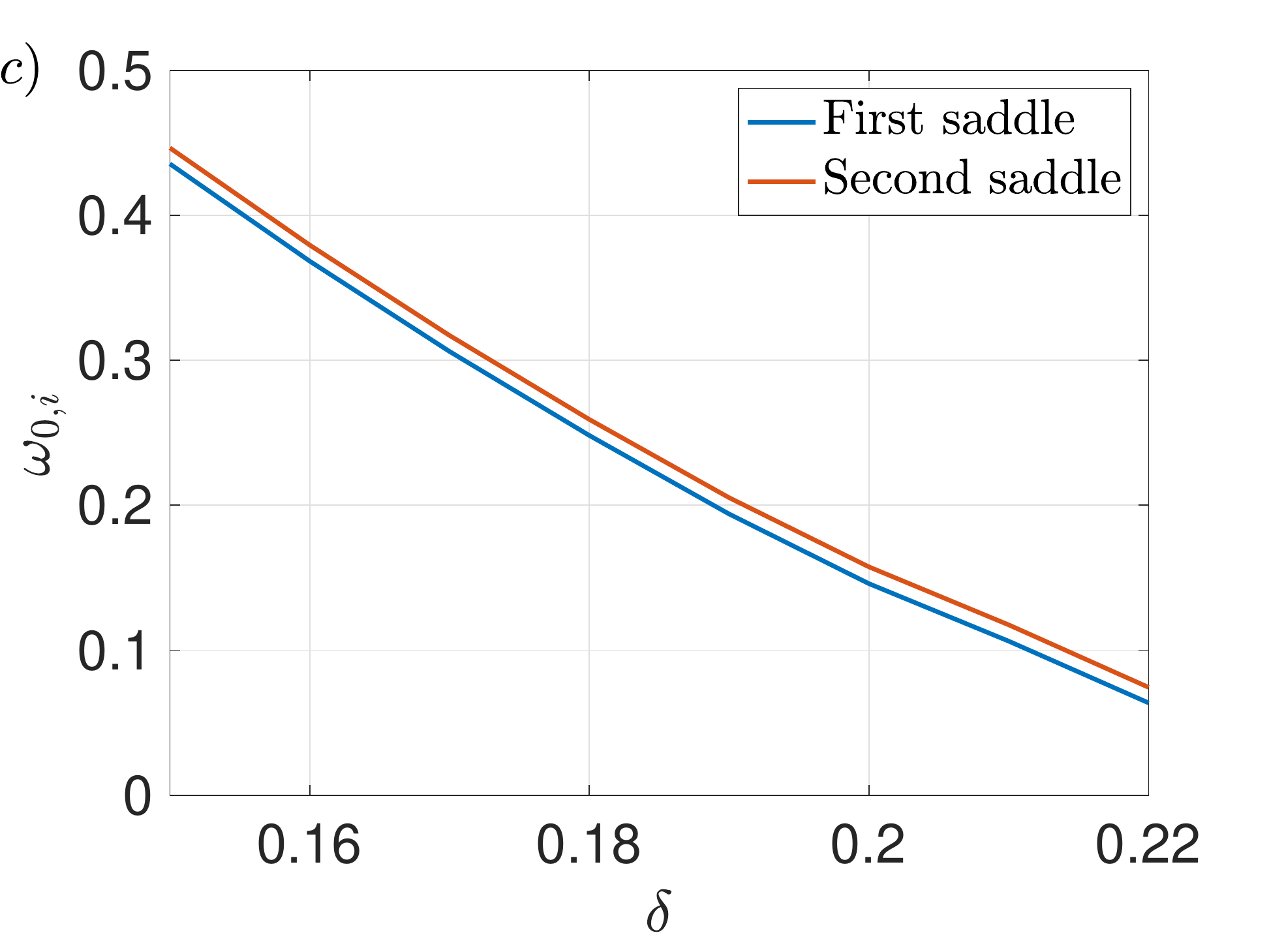}\includegraphics[clip=true, trim= 0 0 0 0, width=0.5\textwidth]{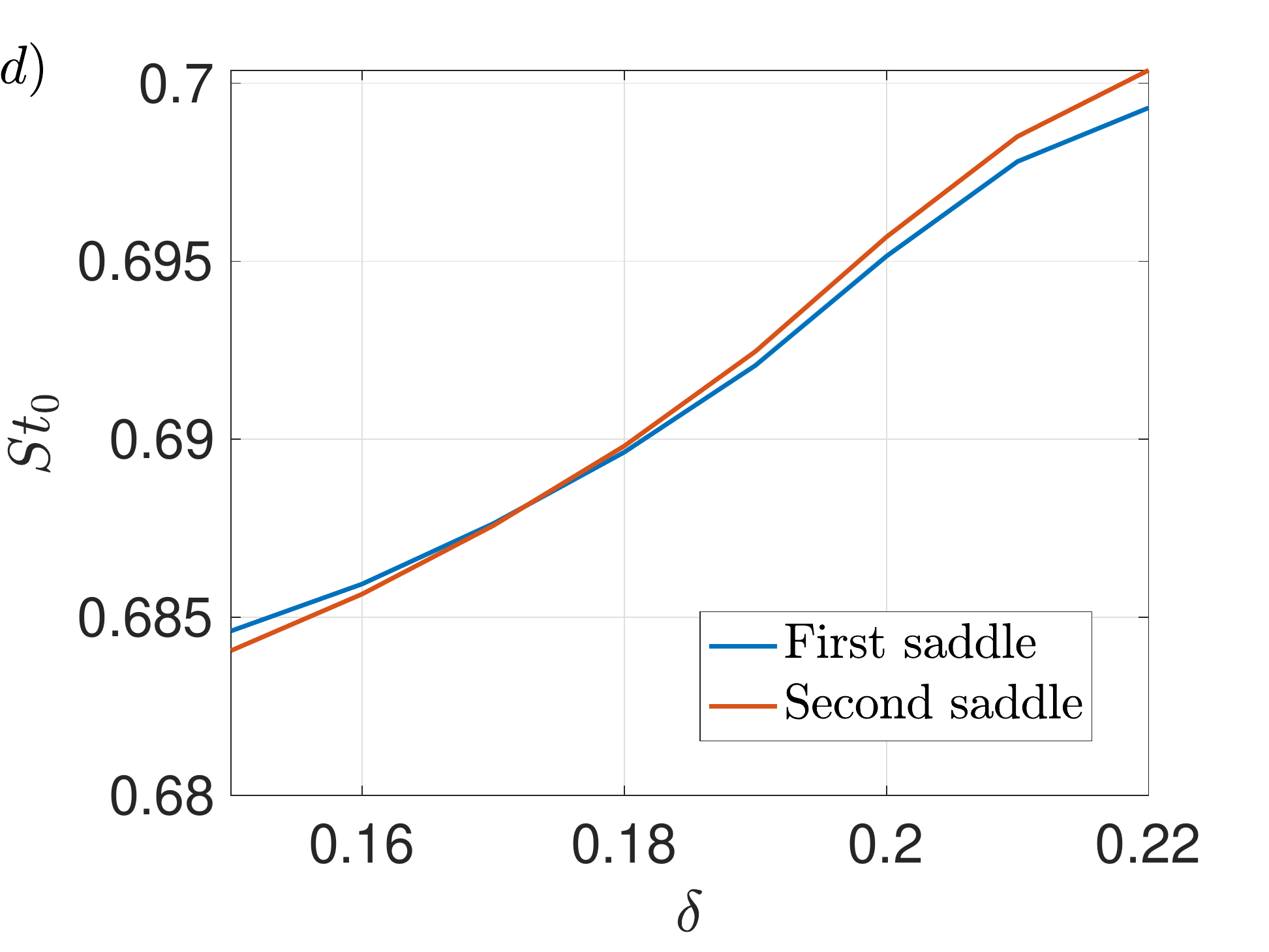}
\caption{Imaginary frequency $\omega_{0,i}$ (a,c) and Strouhal number $St_0$ (b,d) of the first and second saddle points for $NPR=2.1$ and $m=0$ as a function of shock amplitude $A_{sh}$ and shear layer thickness $\delta$. Results for $\delta=0.15$ (a,b) and $A_{sh}=0.02$ (c,d).}
\label{fig:SaddlesParametric}
\end{figure}

The sensitivity of the temporal growth rates and Strouhal number to the shock amplitude is shown in figure \ref{fig:SaddlesParametric}(a,b). As suggested by the results in figures \ref{fig:SaddlesAll}(a,c), the difference in frequency between the two saddles is small for $A_{sh}=0.02$; figures \ref{fig:SaddlesParametric}(a,b) show that this difference decrease as $A_{sh}$ is decreased, and the two saddles coalesce for vanishing values of shock-cell amplitude. No saddle is observed for $A_{sh}=0$, which points to the absence of absolute instability in the locally parallel case. Overall, both temporal growth rates and Strouhal number of the saddle are fairly insensitive to changes in shock-cell amplitude, as shown in figures \ref{fig:SaddlesParametric}(a,b). On the other hand, these quantities are strongly affected by the shear-layer thickness $\delta$. Figure \ref{fig:SaddlesParametric}(c) shows a monotonic decrease in the growth rates of the structures associated with both saddles as $\delta$ increases, which is in line with the decrease in the spatial growth rate of the Kelvin-Helmholtz mode with increase of shear-layer thickness \citep{michalke1984survey}. The Strouhal numbers associated with the saddle increase slightly with the increase of $\delta$, but these values still remain in the frequency range where screech is expected to occur for this value of $NPR$ \citep{mancinelli2019}. 

These results altogether suggest that a pocket of absolute instability, obtained in a periodic stability analysis using velocity profiles from various $x$ stations, might be present in real shock-containing flows, where the jet spreading will have a stabilising effect on the disturbances. This is also similar to the studies of pockets of absolute instability in wakes \citep{chomaz_huerre_redekopp1988}. Still, differently from the wake case, the absolute instability in shock-containing jets is triggered by the periodicity induced by the shock-cells, and no pocket of absolute instability is observed for such jets in the locally parallel framework. We emphasise that, contrary to former models that needed empirical inputs, the absolute instability analysis in the spatially periodic framework predicts the screech frequencies directly as a function of the jet operating condition. Increasing the shear-layer thickness leads to small variations on the predictions within the range of frequencies in which screech is observed. No other input from experiments is needed.


\subsection{Comparison of absolute instability eigenfunctions with data}

\begin{figure}
\centering
\includegraphics[clip=true, trim= 50 20 20 10, width=\textwidth]{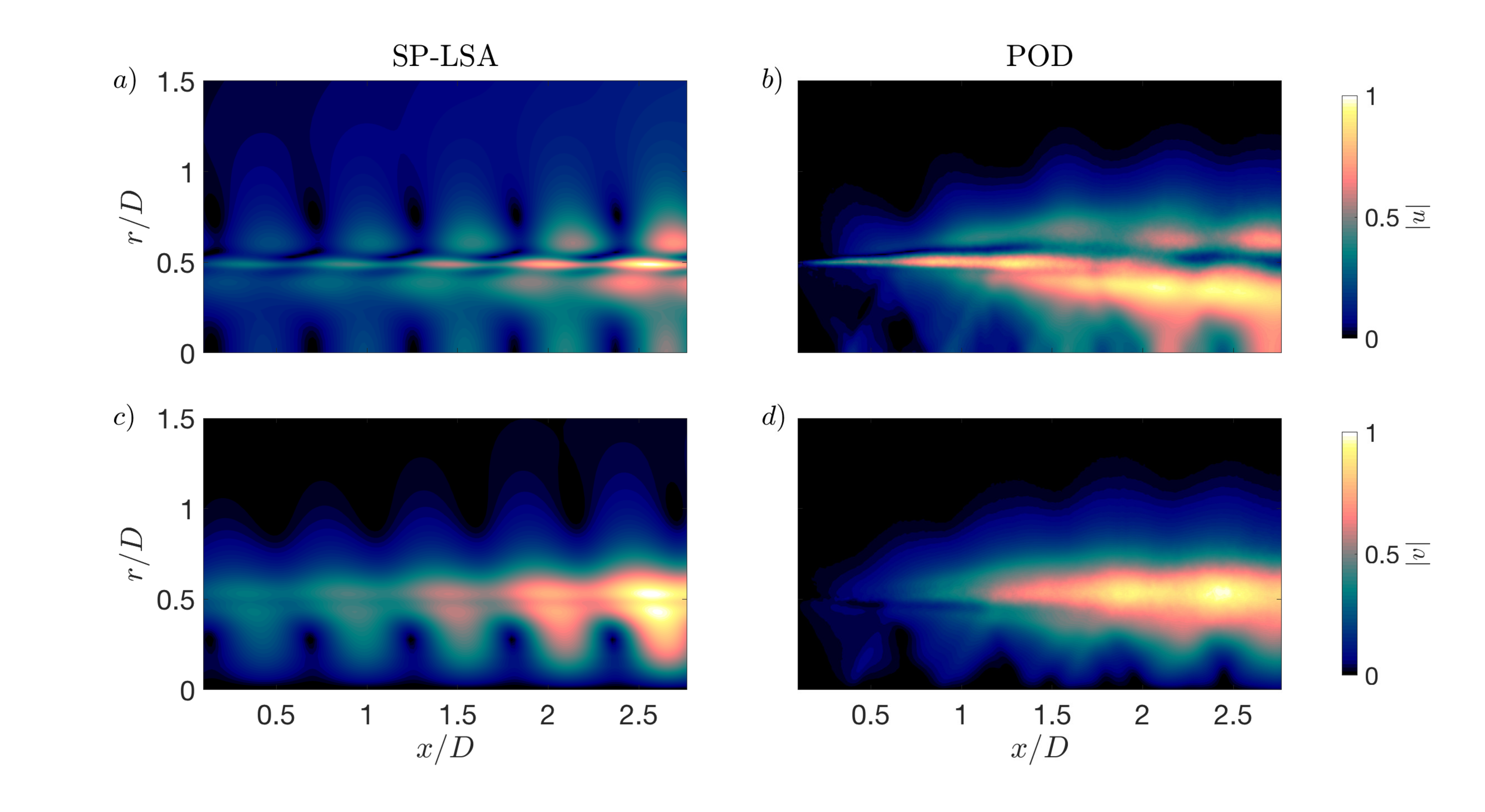}
\caption{Comparison between the shapes of the modes close to the saddle and POD modes for $NPR=2.1$ and $m=0$. The spatial growth rates of the modes ($\mu_i$) are now included in the reconstruction. Absolute value of streamwise and radial velocities are shown in (a) and (c) for the spatially periodic stability analysis, and streamwise and lateral velocities of the POD modes ($NPR=2.1$) reported in \cite{EdgingtonMitchell2020} are shown in (b) and (d). All modes are normalised by their maximum.}
\label{fig:ModesSaddlem0}
\end{figure}

Further analysis of the structure of the modes close to the the resonant condition may confirm if the absolute instability mechanism is actually representative of the screech phenomenon. Here, we wish to discern if the large-time response to an impulse applied to a spatially-periodic shock cell pattern resembles the dominant coherent structures in a screeching jet. The axisymmetric modes for $NPR=2.1$, $St = 0.6841$ and $\omega_i=0.4465$ are presented in figure \ref{fig:ModesSaddlem0}(a,c), where the absolute value of streamwise and radial velocities are shown, respectively. The spatial growth of SP-LSA modes is also included in the reconstructions. Comparing figures \ref{fig:ModesSaddlem0}(a,c) to figures \ref{fig:absUmodxrm0}(a-d), it is clear that the effect of the modulation by the shock-cell structure is minor close to the resonant condition (which may be related to the spatial distribution of the modes further downstream in figure \ref{fig:Umodcentre21}(b)), and the interference pattern between the two waves involved in the saddle is now dominant. This interaction leads to strong modulation of the resulting wavepacket structure throughout the domain \citep{EdgingtonMitchell2020}. The structure of the modes close to the saddle can also be compared to the most energetic coherent structure coming from a POD of experimental data, presented in \cite{edgington-mitchell_jaunet_jordan_towne_soria_honnery_2018} and \cite{EdgingtonMitchell2020}, shown in figures \ref{fig:ModesSaddlem0}(b,d). Even though the modes from the spatially periodic analysis do not capture the growth/decay behaviour of the wavepacket (which is due to the increasing shear-layer thickness, not included in the model), the overall modulation pattern is quite well reproduced by the modes. Such agreement can be especially observed close to the centreline and outside the shear layer, where a wavy pattern is clearly seen. The POD modes are also more radially spread across the shear layer and potential core, which is also likely due to the jet spreading.

\begin{figure}
\centering
\includegraphics[clip=true, trim= 0 60 0 100, width=\textwidth]{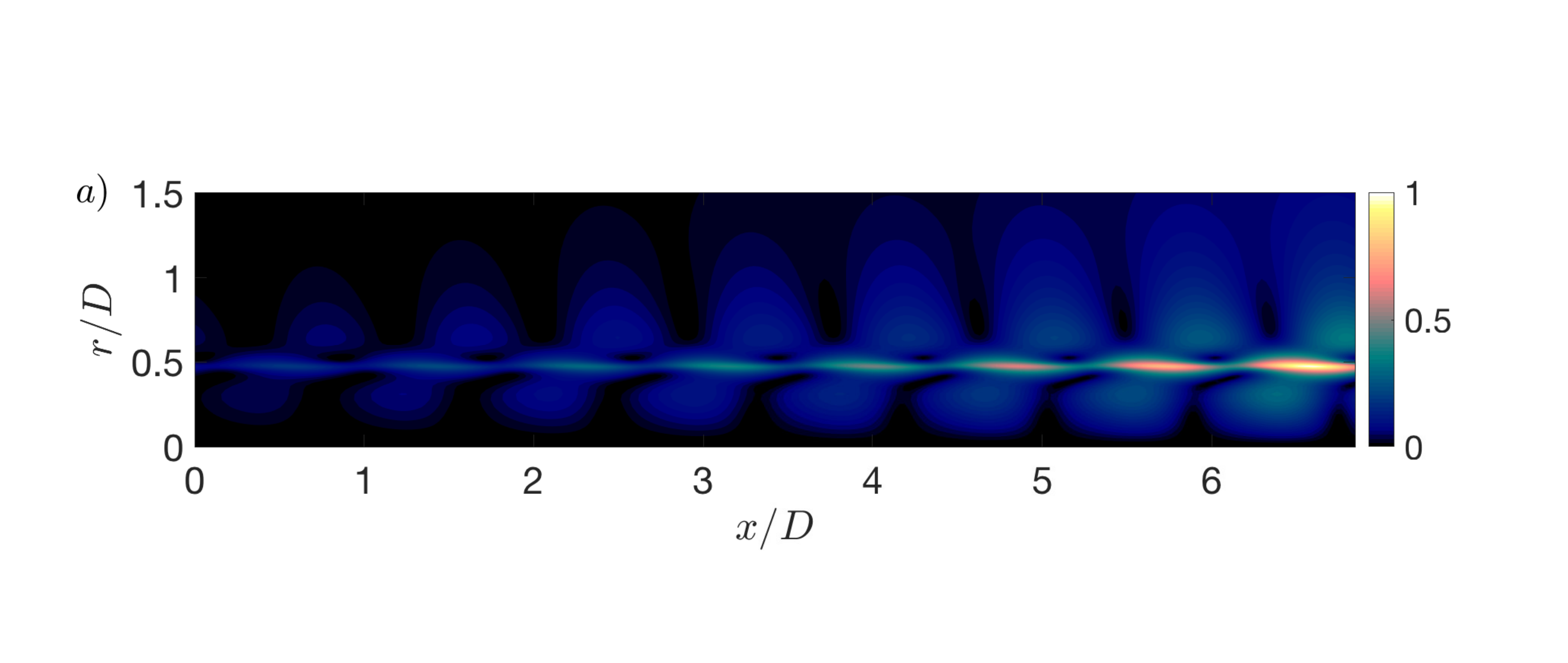}
\includegraphics[clip=true, trim= 0 80 0 120, width=\textwidth]{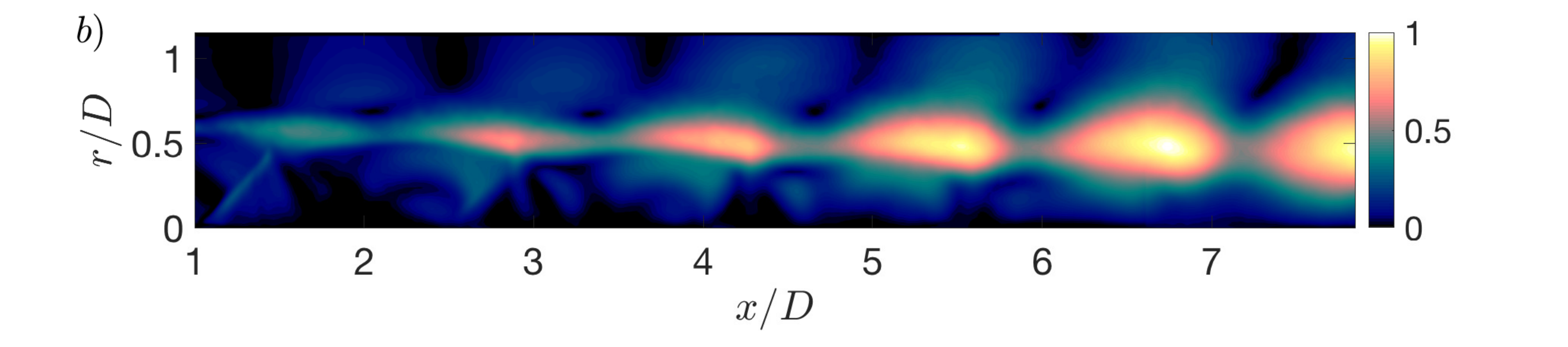}
\includegraphics[clip=true, trim= 0 60 0 100, width=\textwidth]{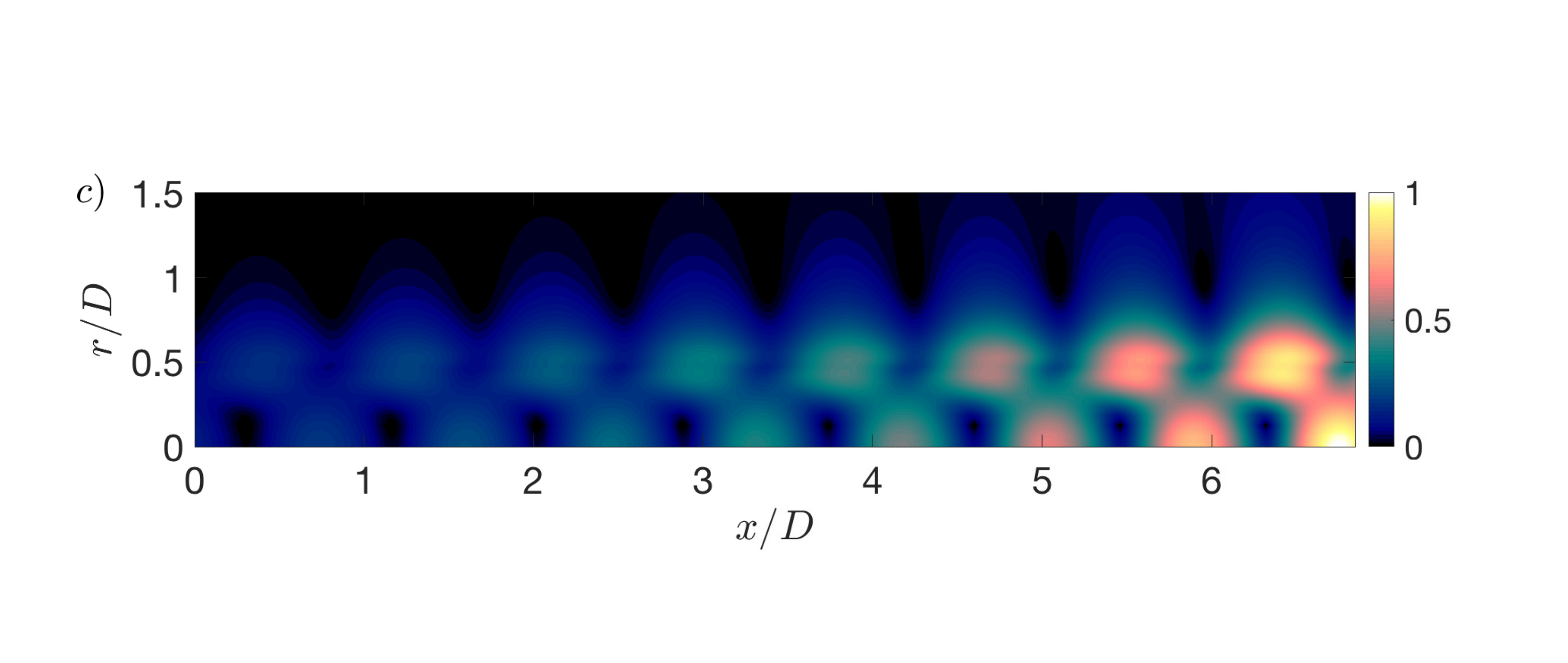}
\includegraphics[clip=true, trim= 0 80 0 120, width=\textwidth]{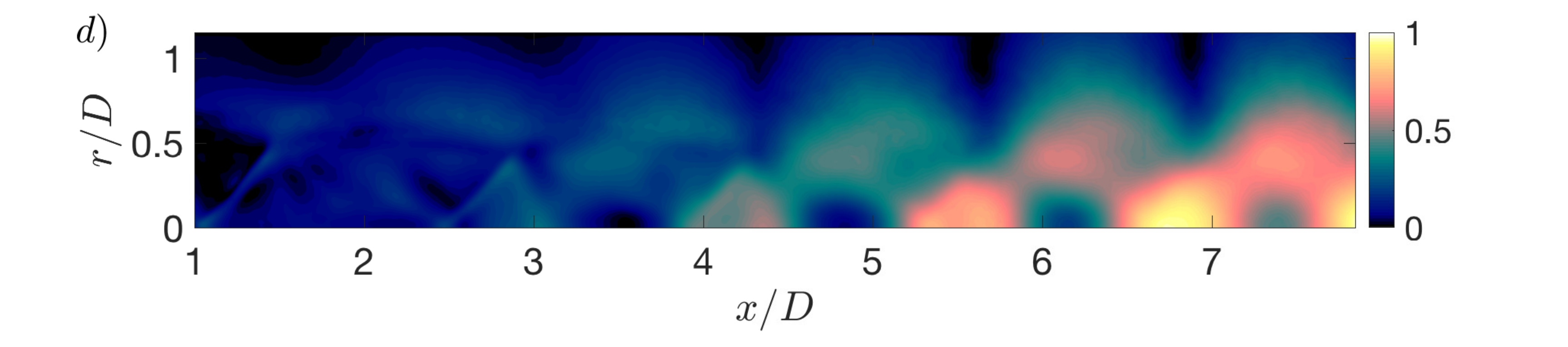}
\caption{Comparison between the shapes of the modes close to the saddle and POD modes for $NPR=2.4$ and $m=1$. The spatial growth rates of the modes ($\mu_i$) are now included in the reconstruction. Absolute value of streamwise and radial velocities are shown in (a) and (c) for the spatially periodic stability analysis, and streamwise and lateral velocities of the POD modes ($NPR=3.4$) reported in \cite{EdgingtonMitchell2020} are shown in (b) and (d). All modes are normalised by their maximum.}
\label{fig:ModesSaddlem1}
\end{figure}

The spatial reconstruction of the $m=1$ mode for $NPR=2.4$, $St = 0.448$ and $\omega_i=0.204$ is shown in figures \ref{fig:ModesSaddlem1}(a,c). As in the axisymmetric case, the interference pattern formed by the interaction between the two waves is much stronger than the modulation by the shock-cell structure alone. Compared to the non-resonant condition, changes in the mode structure are also seen over the entire mode, especially for the radial velocity; in this velocity component, an alternation of peak velocities around the shear-layer and centreline is observed as consequence of the regions of the peak of each wave (as shown in figures \ref{fig:absUmodxrm1}(b,d)) and the interference pattern. The only modal decomposition data available for this operating condition are the DMD modes presented by \cite{li_zhang_hao_he_2020}, where only the real part of the modes are shown, hindering the identification of the modulation. Thus, a comparison with modes from another operating condition ($NPR=3.4$, studied in \cite{edgington-mitchell_oberleithner_honnery_soria_2014} and \cite{EdgingtonMitchell2020}) is performed here. Even though the flow for this higher Mach number case is slightly more complicated (especially considering the presence of a small Mach disk) it is also dominated by an $m=1$ screeching mode, as in the present case. It is also worth highlighting that the shock-cell wavelength for this higher $NPR$ case is larger, so the modulation wavenumber of the mode close to the saddle will differ from the experimental result; thus, only qualitative comparisons about the structure of the resonant mode may be performed in this case. The absolute value of the streamwise and radial velocities for this case are shown in figures \ref{fig:ModesSaddlem1}(b,d). As in the axisymmetric case, a good agreement between the structure of the spatially periodic modes and the POD modes is obtained for $m=1$ disturbances. The wave-interaction pattern follows the experimental data quite closely, especially in the outer region of the jet, for the streamwise velocity. For this component, the viscous effects and the shear-layer development also lead to modes more radially spread. A remarkable agreement for the radial velocity is observed, with both fields following the particular peak-shift behaviour between the centreline and the shear-layer.

\begin{figure}
\centering
\includegraphics[clip=true, trim= 0 50 0 0, width=\textwidth]{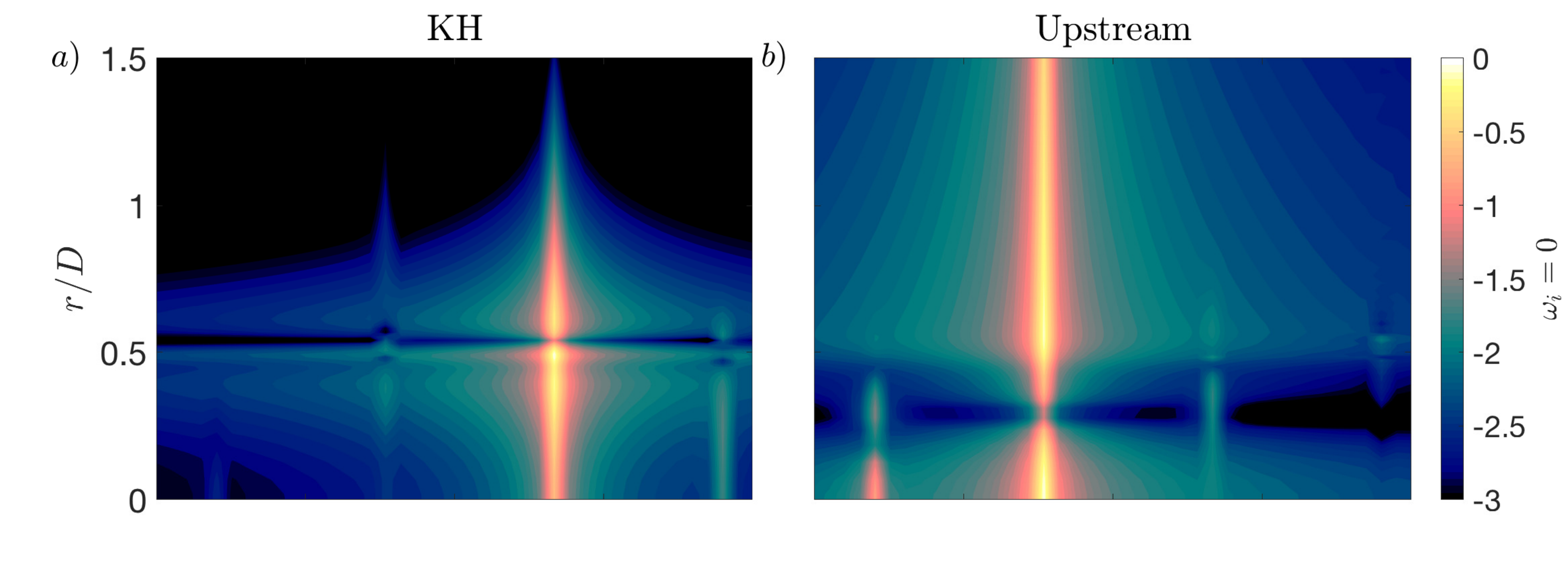}
\includegraphics[clip=true, trim= 0 50 0 0, width=\textwidth]{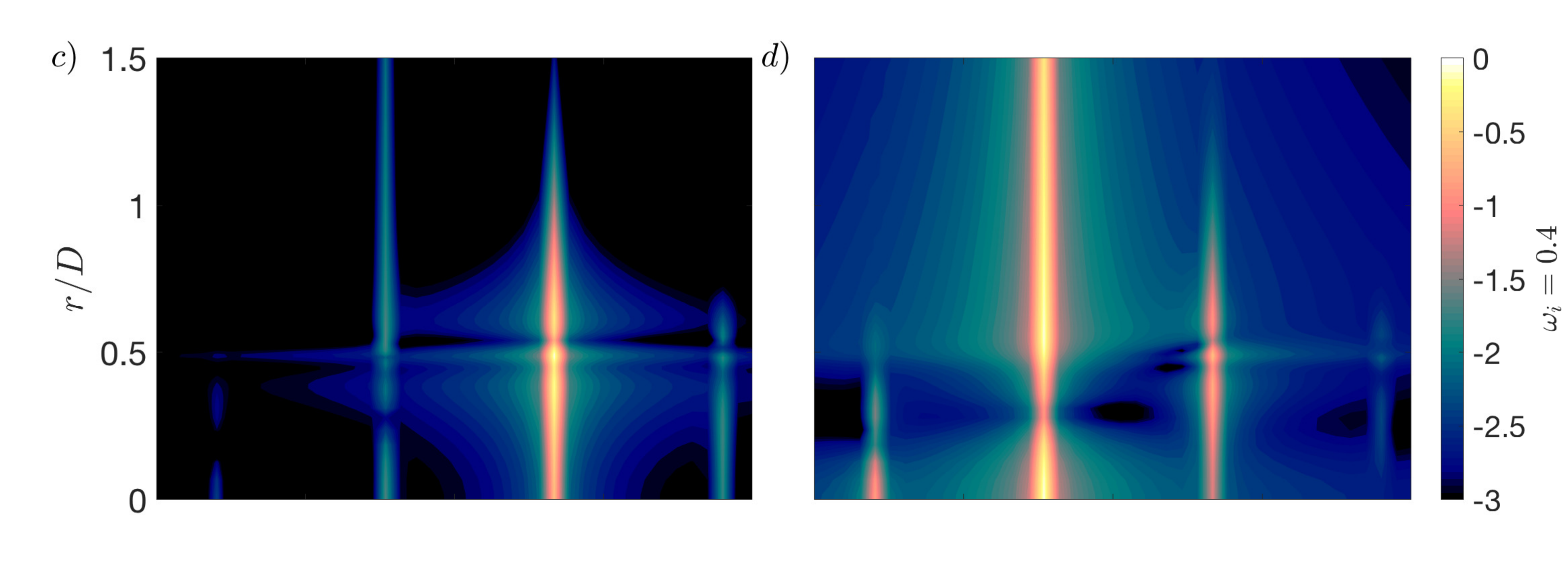}
\includegraphics[clip=true, trim= 0 0 0 0, width=\textwidth]{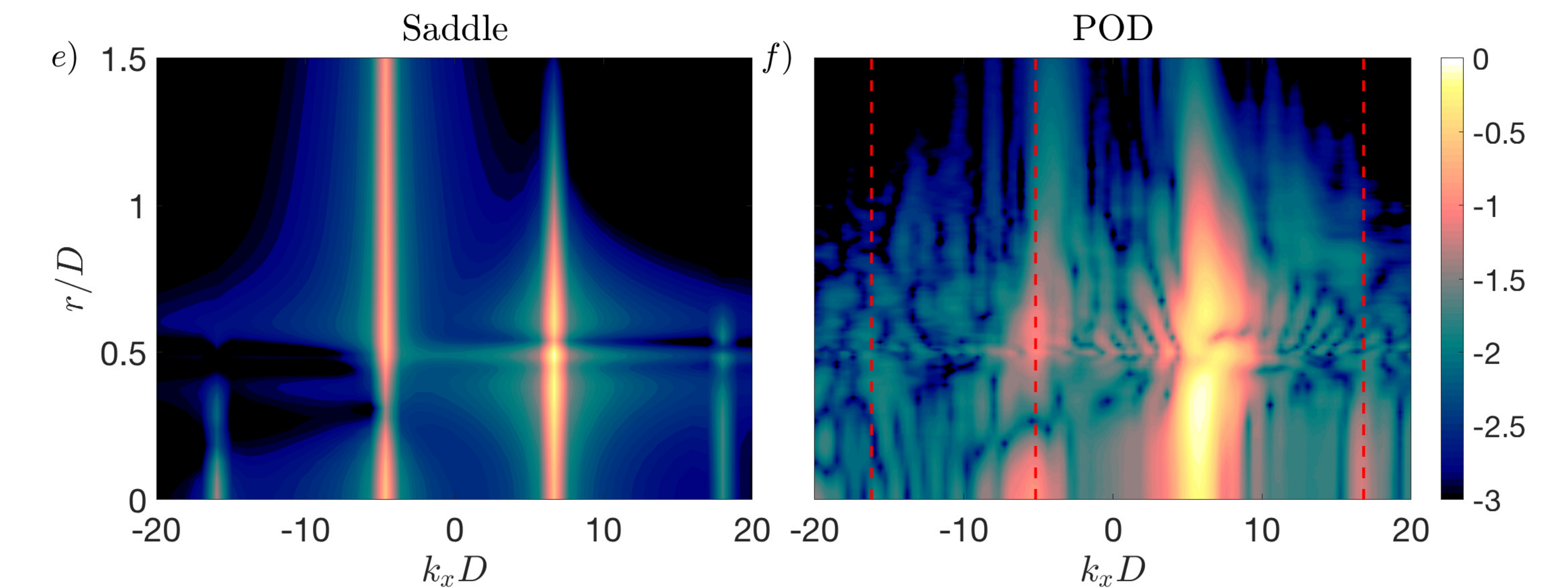}
\caption{Wavenumber spectrum related to the spatial reconstruction of the relevant eigenmodes (neglecting $\mu_i$), for $NPR=2.1$ and $m=0$. Spectra for Kelvin-Helmholtz (a,c) and upstream travelling modes (b,d) are shown for $\omega_i=0, 0.4$ and $St=0.6846$. Spectrum of the mode close to the second saddle point ($\omega_{0,i}=0.446$, $St_0=0.6841$) is shown in (e). The same spectrum for the $NPR=2.1$ POD mode, from \cite{EdgingtonMitchell2020}, is also reproduced (f), with the modulation wavenumbers related to the leading KH wavenumber are depicted by the red lines. All plots are normalised by their maximum, and the magnitudes are presented in log scale.}
\label{fig:WavenSpecm0}
\end{figure}

The presence of both KH and upstream waves in the mode at the saddle point can be more clearly observed in the analysis of the wavenumber spectrum of this mode. The streamwise velocity of the modes are reconstructed in the first ten shock-cells ignoring the spatial growth rate ($\mu_i$), in order to isolate the oscillatory behaviour of the modes. It is desirable to evaluate the streamwise wavenumbers present in the eigenfunctions of each wave as one approaches the saddle; thus, a spatial Fourier transform of the reconstructed mode is performed for different values of $\omega_i$. This analysis is equivalent to the one performed by \cite{EdgingtonMitchell2020} on POD and global stability modes. Considering that only wavenumbers $\mu_r+Nk_{sh}$ are allowed in each mode, the spatial spectra provides a good estimate of the relative energy of the different wavenumbers included in each mode, allowing for a clearer identification of the different waves at the saddle point. Figures \ref{fig:WavenSpecm0}(a-d) show the wavenumber spectra of the modes related to the KH (a,c) and guided jet (b,d) waves for $\omega_i=0, 0.4$. These plots show the energy content of each wavenumber of the reconstructed mode using a fast Fourier transform (\textit{fft}). The discrete wavenumbers allowed in the eigenfunctions will appear as most energetic; energy in wavenumbers other than $\mu_r+Nk_{sh}$ must be disregarded, as it is related to contour interpolation and to the truncation of the domain for the \textit{fft}, as the wavenumbers $\lambda_{sh}$ and $2\pi/\mu_r$ are not multiples of each other. For zero imaginary frequency, both modes have peaks at the expected wavenumbers (positive for the KH, negative subsonic for the guided jet mode), and small energy peaks related to the modulation analysed in section \ref{sec:modulation} are also observed. As the imaginary frequency is increased to $\omega_i=0.4$, the energy of wavenumbers related to upstream waves start to increase in the KH mode (and equivalently in the upstream wave), until both modes become one at the saddle point. There, it is impossible to separate both waves, and the resulting mode has energy peaks at the wavenumbers of both waves, as shown in figure \ref{fig:WavenSpecm0}(e). At this position, both modes share the same eigenvalue $\mu$, and only wavenumbers $k=\mu_r+N k_{sh}$ are allowed in the eigenfunction. Hence, the wavenumbers of both waves that compose the saddle must observe the relationship $k_{kh}-k_{sh}=k_{upstream}$, where $k_{kh},k_{upstream}$ are the peak wavenumbers of the KH and upstream waves (associated with $N=0,-1$, respectively), identified by a continuation of the wavenumber spectra from $\omega_i=0$ to $\omega_i=\omega_{0,i}$. This is exactly the wavenumber relation derived by \cite{TamTanna1982}, and confirmed by \cite{EdgingtonMitchell2020}. In fact, comparing the wavenumber spectrum of the mode close to the saddle point and the POD mode for the same Mach number, shown in figure \ref{fig:WavenSpecm0}(f) (modulation wavenumbers indicated by the red dashed lines), the same overall structure is observed. The most energetic wavenumbers in both analyses are found around the same positions, which provides a further indication that the absolute instability is the phenomenon at play in screech.

The correspondence between the modes related to the absolute instability and the most energetic modes in screeching jets, together with the accurate prediction of the screech frequencies by the present analysis, strongly suggests that screech is actually an absolute instability mechanism, or at least that such mechanism is triggering the resonance phenomenon. For the cases studied herein, the saddle that characterises this phenomenon is formed between the Kelvin-Helmholtz mode and the discrete guided-jet mode, supporting the hypothesis raised by \cite{gojon2018aiaa} and \cite{edgington-mitchell_jaunet_jordan_towne_soria_honnery_2018} that this upstream mode is responsible for closing the resonance loop. This also shows that, even though the presence of shocks in the jet changes the overall characteristics of the flow quite abruptly, the main changes in the dynamics of the jet are due to the formation of a periodic structure, at least in cases where the position of a downstream reflection point is not clear, which is the case for free jets.

\begin{figure}
\centering
\includegraphics[clip=true, trim= 0 0 0 0, width=0.8\textwidth]{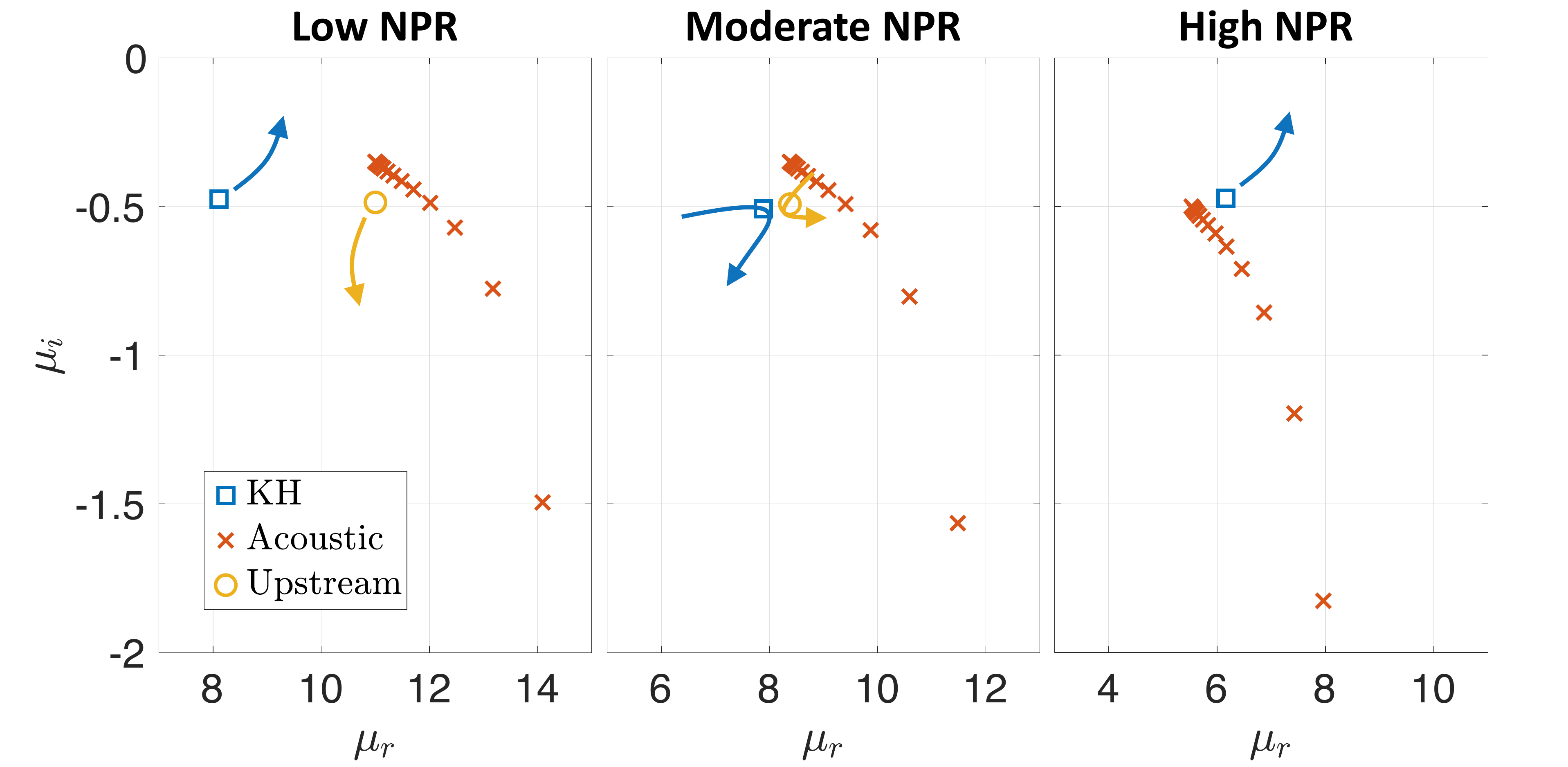}
\caption{Sketch of the cut-on/cut-off mechanism of screech mode A1.}
\label{fig:PSD_Sketch}
\end{figure}

The acoustic spectrum of a screeching jet as a function of Mach number (as presented by \cite{mancinelli2019}, for instance) can now be analysed in light of the present results. For very-low supersonic cases (when screech starts to occur), the shock-cell wavenumber $k_{sh}$ is too high, and the modes are too far apart in the spectrum to interact. As the Mach number is increased, $k_{sh}$ decreases, and both modes are now in the same region, allowing for the occurrence of a saddle point between axisymmetric KH and upstream modes. Increasing $M_j$ leads to a further decrease in the shock-cell wavenumber; in this case, the Kelvin-Helmholtz mode crosses the acoustic branch before the guided jet mode becomes cut-on, and no saddle is found (although the presence of secondary wavenumbers in the shock-cell structure may provide alternative paths for interaction \citep{Nogueira2020A1A2}). Still, this decrease in the value of $k_{sh}$ will allow for the interaction between these waves for $m=1$, giving rise to the scar-like plot for the B mode. The cut-on/cut-off process of screech mode A1 is summarised in figure \ref{fig:PSD_Sketch}. Currently, no reason for the switch between B-C-D modes is provided by the model, and such analysis is outside of the scope of this work.

\section{Relationship with previous models and impact on the jet dynamics}
\label{sec:relationship}

We can now focus on providing a link between the present results and some of the models previously developed in the literature for predicting screech. The equivalency between the original formulation of \cite{powell1953} and formulations based on the standing wave \citep{panda1999} and shock-wavepacket interaction \citep{TamTanna1982} was derived by \cite{EdgingtonMitchell2019}; essentially, all these formulations relate the phase velocity of the Kelvin-Helmholtz mode with the wavenumber of acoustic waves using some sort of phase argument based on the shock-cell spacing, or some other characteristic length. Considering that the phase velocity of the KH mode is well captured by experiments and linear models (such as the vortex-sheet), and that the wavenumber of the guided jet mode is usually very close to the wavenumber of upstream-travelling acoustic waves for a given frequency, these formulations usually manage to approximate the main ingredients of the screech mechanism using a single real-valued equation. Among these previous approaches, the one that most closely resembles the absolute instability mechanism is the one proposed by \cite{TamTanna1982}. By considering the interaction between wavepackets and the shock-cell structure, the authors proposed that the presence of the shocks would energise the wavenumbers $k=k_{kh} \pm k_{sh}$. The authors were already considering the periodicity of the spectrum shown in figure \ref{fig:EigsNPR21Ash0}, even though they did not consider the relevance of the upstream-travelling mode in the resonance closure mechanism. Recently, an alternative formulation using the same wavenumber relation \citep{Nogueira2020A1A2} was used in association with linear stability analysis and the guided jet mode to predict screech frequencies, leading to accurate results; these wavenumber relations can be seen as a first approximation of the position of the saddle point in the present formulation. 

It is now well known that screech is associated with a global instability of the jet (see \cite{BENEDDINE2015}, for example), and the present analysis offers an explanation of how this global instability is triggered based on the spatial periodicity of the flow. An alternative formulation based on both phase and magnitude of the resonance loop was proposed by \cite{mancinelli2019,mancinelli2020}, following the formulation of \cite{jordan_jaunet_towne_cavalieri_colonius_schmidt_agarwal_2018}. The authors proposed that screech could be explained in the locally parallel framework as a long-range feedback using a complex resonance model, which depends on the wavenumbers of the different waves involved in the phenomenon, a characteristic length (usually chosen as the position of one of the shock-cells), and reflection coefficients (related to the amount of energy transferred to upstream waves when large-scale structures travel past a shock). The present analysis proposes a somewhat different mechanism; in this case, screech is seen as an absolute instability mechanism defined by the periodicity of the flow, by the bands of existence of the guided jet mode, and by the characteristics of the Kelvin-Helmholtz mode. Here, the effect of the nozzle is restricted to its secondary impact on the jet plume, and no upstream/downstream reflections are directly considered; the jet is absolutely unstable simply due to the periodicity and the wavenumbers of the relevant modes. This is similar to the effect of back-flow in the absolute instability of wakes, which may lead to resonance even in the absence of a bluff body that generates the wake  \citep{MONKEWITZ1987,huerre1990local,pier_huerre_2001}.

Even though resonance in the spatially periodic model may be closed without the need for a nozzle, this mechanism may be modified by the presence of an upstream reflection condition. As observed in previous experiments \citep{PONTON1992531,raman_1997,RAMAN1998}, the amplitude of the screech tone can be strongly affected by the nozzle lip thickness, while its frequency usually remains unchanged. This is in line with the present results, where resonant frequencies are determined by the real part of the frequency of the saddle point $\omega_{0,r}$. Since only screech amplitude is affected by the nozzle, it is conjectured that the nozzle will have an effect on the amplification of the resonant mode $\omega_{0,i}$, in the global framework. Consideration of non-periodicity in the flow (such as in the region very close to the nozzle) will lead to a decoupling between upstream- and downstream-travelling waves generated by the absolute instability. This can generate reflected waves that could constitute an additional source of amplification of the resonant mode. In this sense, the reflection by the nozzle can still be important in leading to a sufficiently strong global instability for a high amplitude tone.

That the output of the present model is generally insensitive to the shock-cell amplitude is perhaps a somewhat counter-intuitive result, but it is consistent with both prior models and experimental results for screeching jets. In the shock-leakage model of \cite{manning1998numerical,manning2000numerical}, the same qualitative behaviour was observed whether the incident wave was a strong shock, a near-isentropic compression wave, or a Gaussian wave. The latter extension of the model by \cite{ShariffManning2013} demonstrated that the ``leakage'' behaviour could be preserved even when shock amplitude was entirely neglected as a parameter in the model. Experimental studies by \cite{raman_1997} and \cite{RAMAN1999543} have also suggested that while shocks are required for screech, there is little correlation between shock strength and screech tone amplitude.

The present results also shed light on the analysis performed by \cite{EdgingtonMitchell2020}. Using experimental data and global modes, the authors showed the existence of two different waves resulting from the shock-wavepacket interaction in a screeching mode: the guided jet mode and the soft-duct mode. They also showed that the superposition between the KH mode and these other waves is quite strong in the flow field, even though this effect could not be separated from the modulation caused by the shock-cells. Here, we confirm these previous results, showing that the presence of both KH and guided jet modes in the flow field is a direct consequence of the absolute instability mechanism. We also show that the modulation due to the interaction between the two waves is much stronger than the modulation given by the shocks (compare figures \ref{fig:absUmodxrm0} and \ref{fig:ModesSaddlem0}, for instance), supporting this previous analysis. The importance of the soft-duct mode could not be assessed in the absolute instability framework (since the saddle must be formed between upstream- and downstream-travelling modes).

Some remarks about the impact of the present findings on the flow dynamics must be provided; this will be analysed following the interpretation of convective and absolute instabilities developed by \cite{huerre1990local}. In convectively unstable flow, such as ideally expanded jets, disturbances are generated by some sort of forcing upstream (where the flow is most sensitive) and convected away from the source; in the case of absolutely unstable flows, disturbances spread in both downstream and upstream directions, and contaminate the entire flow field. For that reason, convectively unstable flows can be seen as noise amplifiers, while absolutely unstable flows behave as oscillators; in the latter case, self-sustained resonant states of the flow may be observed. This is in line with the original observations of \cite{powell1953}, who also pointed out the presence of upstream-travelling waves in the flow, connected to the development of the large-scale vortical structures. It is also aligned with the classical view that screech is a self-sustaining resonant phenomenon. Thus, the basis of the overall interpretation of screech as a physical phenomenon is still applicable for the present case.

It is also important to highlight the general applicability of the results, and provide some caveats concerning the application to spatially developing flows. First, considering that screech is observed in all sorts of shock-containing jets (see \cite{EdgingtonMitchell2019}), and that the basic characteristics of the phenomenon is shared by free jets with different geometries, showing that this phenomenon is actually a consequence of an absolute instability mechanism also impacts the interpretation of screech in these other cases. Thus, screech in rectangular, elliptical and twin jets could also be considered as an absolute instability phenomenon, when the equivalent modes are considered in each geometry. Still, for all these cases, one should be aware that the spatial development of the flow may have an impact on the analysis. In practice, one is expected to find regions of the flow in which such a mechanism is at play. The effect of increasing shear layer thickness further downstream may lead to flow to switch back to a convectively unstable state, and eventually to a stable state. Thus, the present model can be seen as an approximation of the overall behaviour of the flow -- but one that captures some of the key features of the flow in the resonating condition. 

\section{Conclusions}
\label{sec:concl}

This work focuses on the application of spatial linear stability analysis to flows with spatial periodicity, such as shock-containing jets. In order to account for the periodicity, the Floquet ansatz is used to write the general form of the disturbances, and streamwise Fourier modes are used to account for the flow periodicity \citep{Brevdo1996,Herbert1988}. Solutions of the linearised Navier-Stokes system written as an eigenvalue problem are obtained numerically using standard methods. This formulation has a direct connection with the locally parallel case: when the amplitude of the periodic structure is zero, the locally parallel results are recovered with the periodicity of the flow. This allows for a clearer categorisation of the different modes based on the locally parallel analysis. It is shown that small shock-cell amplitudes do not lead to significant changes in the eigenvalues, but the eigenmodes are significantly modified. Using the spatially periodic linear stability analysis, the modulation of the different waves involved in screech could be obtained directly as function of frequency, Mach number and shock-cell amplitude, with a computational cost similar to the locally parallel analysis. This tool opens new pathways to analyse shock-containing jets, where the effect of a spatially periodic shock-cell structure is directly accounted for in the model, resulting in modes that can be directly compared with experiments in a straightforward manner.

One of the main effects of including the shock-cell structure in the linear stability analysis is related to a change in the nature of the instabilities supported by the flow. It is well known that cold jets are subject to a convective instability mechanism, which gives rise to the Kelvin-Helmholtz mode. By including the periodicity in the analysis, we show that cold jets can also be subject to an absolute instability mechanism, leading to disturbances that spread both downstream and upstream. Thus, a mechanism for the appearance of a globally unstable mode in shock-containing jets (as in \cite{BENEDDINE2015}) is brought to light. The saddle point analysis shows that the absolute instability is underpinned by an interaction between the Kelvin-Helmholtz mode and the guided jet mode, supporting the theory that this discrete upstream mode is responsible for closing the resonance phenomenon in screeching jets \citep{gojon2018aiaa,edgington-mitchell_jaunet_jordan_towne_soria_honnery_2018} for both $m=0,1$ cases. Frequencies and modes related to the saddle points are markedly similar to the ones obtained by experiments and simulations, for both axisymmetric and helical disturbances, corroborating the connection between the present results and the physical mechanism behind screech. 

Overall, the present analysis provides a mechanistic interpretation of screech based on linear models. In particular, the analysis shows that the reflection at the nozzle is not a necessary ingredient in the selection of the screech frequency, even though the amplification of the resonant mode may still be impacted by the nozzle geometry. Knowing which factors dictate the absolute instability mechanism may not only enhance the understanding of the phenomenon, but also help to design control strategies to suppress the screech tones. In particular, modifications of the mean flow that affect the frequencies of existence of the guided jet mode or the characteristics of the Kelvin-Helmholtz mode can be used to decrease the temporal growth rate of the instability, which can cause the screech tone to vanish. An example of such strategy is in the work of \cite{alkislar2005structure}. By introducing an azimuthal non-uniformity (in the form of micro-jets), the authors managed to suppress the tones in screeching twin-jets. Considering that the introduction of streaks in the flow can affect the growth rate of the Kelvin-Helmholtz mode \citep{marant2018influence,lajus_sinha_cavalieri_deschamps_colonius_2019,nogueira_cavalieri_2021}, these previous results can actually be connected to a mitigation of the absolute instability presented herein. This is a simple example of how the present analysis can be applied to experiments and engineering applications.   \\

Acknowledgements: This work was supported by the Australian Research Council through the Discovery Project scheme: DP190102220. \\
Declaration of Interests. The authors report no conflict of interest.

\appendix
\section{Linear operators for spatially periodic linear stability analysis}
\label{app:linops}

Following \cite{towne2016advancements} and the derivation of section \ref{sec:stability}, the operator $\mathbf{L}$ can be decomposed as $\mathbf{L}_c-\ii \omega \mathbf{I}$, where $\mathbf{I}$ is the identity matrix, and $\mathbf{L}_c$ is given by

\begin{eqnarray}
\mathbf{L}_c=
    \setlength{\arraycolsep}{4pt}
    \renewcommand{\arraystretch}{1.3}
    \left[
    \begin{array}{ccccc}
    U_x D_x - \partial_x U_x  &  \partial_x \bar{\nu}-\bar{\nu} D_x  &  \partial_r \bar{\nu}-\bar{\nu}(D_r+\frac{1}{r})  &  -\ii m \frac{\bar{\nu}}{r}  &  0  \\
    \partial_x P  &   U_x D_x + \partial_x U_x  &  \partial_r U_x  &  0  &  \bar{\nu} D_x  \\
    \partial_r P  &  0  &   U_x D_x &  0  &  \bar{\nu} D_r   \\
    0  &  0  &  0 &  U_x D_x & \ii m \frac{\bar{\nu}}{r}   \\
    0  &  \partial_x P + \gamma P D_x  &  \partial_r P + \gamma P D_r + \gamma \frac{P}{r}  &  \ii m \gamma \frac{P}{r}  &   U_x D_x + \gamma \partial_x U_x  \\
    \end{array}  \right] , \nonumber \\
    \
    \label{eqn:ap.L}
\end{eqnarray}
 
\noindent where $D_x$ and $D_r$ are the streamwise and radial differential operators, and $\gamma$ is the specific heat ratio. The operator $\mathbf{L}_{\mu}$ can be written as

\begin{eqnarray}
\mathbf{L}_\mu=-
    \setlength{\arraycolsep}{5pt}
    \renewcommand{\arraystretch}{1.3}
    \left[
    \begin{array}{ccccc}
    \ii U_x &  -\ii \bar{\nu}  &  0  &  0  &  0  \\
    0  &  \ii U_x  &  0  &  0  &  \ii \bar{\nu}  \\
    0  &  0  &  \ii U_x &  0  &  0   \\
    0  &  0  &  0 &  \ii U_x  & 0   \\
    0  &  \ii \gamma P  &  0  &  0  &  \ii U_x  \\
    \end{array}  \right].
    \label{eqn:ap.Lmu}
\end{eqnarray}

\bibliographystyle{jfm}
\bibliography{ref.bib}

\end{document}